\documentclass{aa}
\usepackage{epsfig}
\usepackage{graphicx}
\usepackage{amssymb}
\usepackage{natbib}
\bibpunct{(}{)}{;}{a}{}{,} 

\begin{document}

\title{Quark-nova remnants IV:\\
Application to radio emitting AXP transients}

\author{Rachid Ouyed, Denis Leahy and Brian Niebergal}

\institute{Department of Physics and Astronomy, University of Calgary, 
2500 University Drive NW, Calgary, Alberta, T2N 1N4 Canada\thanks{email:ouyed@phas.ucalgary.ca}}

\date{Received <date>; accepted <date> }

\authorrunning{Ouyed et al.}

\titlerunning{Quark-Nova and radio-emitting AXP transients}

\abstract{
XTE J1810$-$197 and 1E 1547.0$-$5408 are two transient AXPs exhibiting
radio emission with unusual properties. In addition, their spin down rates
during outburst show opposite trends, which so far has no explanation.
 Here, we extend our quark-nova model for AXPs to include transient AXPs, in which the outbursts
  are caused by transient accretion events from a Keplerian (iron-rich) degenerate ring.  
For a ring with
inner and outer radii of $23.5$ km and $26.5$ km, respectively, our model
gives a good fit  to the 
         observed X-ray outburst from XTE J1810$-$197 and the behavior
          of temperature, luminosity, and area of the two X-ray blackbodies with time.
          The two blackbodies in our model are related to 
          a heat front (i.e. Bohm diffusion front)  propagating along the ring's surface   
and an accretion hot spot on the quark star surface.    
              Radio pulsations in our model are caused  by dissipation
    at the light cylinder of magnetic bubbles, produced near the ring during the
    X-ray outburst.  
The delay between X-ray peak
    emission and radio emission in our model is related to the propagation time
    of these bubbles to the light cylinder and  scale with the
    period  as $t_{\rm prop.}\propto P^{\frac{7}{2}-\frac{\alpha}{2}}/\dot{P}^{1/2}$ where
     $\alpha$ defines the radial dependence of matter density in the magnetosphere
      ($\propto r^{-\alpha}$);  for an equatorial wind,
      $\alpha=1$, we predict  a
    $\sim 1$ year and $\sim 1$ month delay for XTE J1810$-$197 and 1E 1547.0$-$5408,
    respectively.  The observed flat spectrum, erratic pulse profile,
     and the pulse duration are all explained in our model as a result of X-point reconnection
     events induced by 
     the  dissipation of the bubbles at the light cylinder.  The spin down
      rate of the central quark star can either increase or decrease depending
       on how the radial drift velocity of the magnetic islands changes with distance
        from the central star.  We suggest an evolutionary connection between
         transient AXPs and typical AXPs in our model. 
}

\maketitle

\section{Introduction}

 Anomalous X-ray pulsars (AXPs) are magnetars with rotation period
  of 2-12 seconds and inferred surface magnetic field strength $B\sim 10^{14-15}$ G (e.g.
  Woods \& Thompson 2006; Kaspi 2007).  In this work we focus
   on 2 AXPs, XTE J1810$-$197 and 1E 1547.0$-$5408, which are the only magnetars known
 to emit in the radio (Camilo et al. 2006). Both are demonstrably 
 transient radio sources, having not been detected in previous
surveys of adequate sensitivity.  
XTE J1810$-$197 is a transient AXP\footnote{In the sense that in
quiescence their surface temperature are as low as those of
 some ordinary  young neutron stars} first detected when
 its X-ray flux increased $\sim 100$-fold compared to a quiescent
  level it maintained for at least 24 years (Ibrahim et al. 2004). 
  Discovered with the {\it Einstein X-ray} satellite in 1980,
    1E 1547.0$-$5408 was eventually identified as a magnetar candidate
     (Gelfand \& Gaensler 2007) with spectral characteristics of an AXP.
In this paper we look at these sources
   in the Quark-Nova context (hereafter QN; Ouyed et al. 2002) building 
    on three previous papers where we explore its application
    to Soft Gamma-ray Repeaters (SGRs) (Ouyed, Leahy, \& Niebergal 2007a; OLNI),
     to AXPs (Ouyed, Leahy, \& Niebergal 2007b; OLNII),
      and to Rotating Radio Transients (RRATs) (Ouyed et al. 2009; OLNIII),
       and superluminous supernovae (Leahy\&Ouyed 2008). But first, we briefly describe
    their observed X-ray and radio properties, during quiescence and bursting phases.

\subsection{The X-ray emission}

In the pre-burst era, XTE J1810$-$197's ROSAT spectrum showed a single blackbody (BB) 
 with temperature $T = 0.18$ keV, an emitting area of $\sim 520\ {\rm km}^2  (d/3.3\ {\rm kpc})^2$,
  and a luminosity of $L_{\rm BB}\sim 5.6\times 10^{33}\ {\rm erg\ s}^{-1} (d/3.3\ {\rm kpc})^2$.
   During its bursting phase,   XTE J1810$-$197  showed a hot blackbody ($T\sim 0.65$ keV) 
with an exponential decay in X-ray luminosity of $\sim 280$ days, 
as well as a warm blackbody ($T\sim 0.3$ keV) decaying at a rate of $\sim 870$ days 
(Gotthelf \& Halpern 2007). 
For the case of 1E 1547.0$-$5408, after its radio detection (Camilo et al. 2007a), 
an X-ray outburst was confirmed
        (Halpern et al.  2008) with a record high luminosity of $\sim
         1.7\times 10^{35}~(d/9.9\ {\rm kpc})^2$ erg s$^{-1}$ and with a total outburst energy
         of $10^{42}\ {\rm erg}  < E_{\rm b} < 10^{43}$ erg.

\subsection{The radio emission}

For XTE J1810$-$197, the radio emission began within 1 yr of  its only known X-ray
    outburst (Camilo et al. 2006 and references therein). At its observed peak
      more than 3 yr after the X-ray outburst, the radio flux density was more
       than 50 times the pre-burst upper limit. 
The X-ray flux has since returned to its quiescent level nearly 4 yrs after the burst.
1E 1547.0$-$5408, although not as well
         sampled as XTE J1810$-$197, exhibits similar variations in flux density and
         was reported with a factor
         of 16 times the pre-burst upper limit (Camilo et al 2007a).

   Trends in radio emission between the 2 sources can be
   summarized as follows:

         \begin{itemize}
         \item  Both are very highly linearly polarized showing  
          a  flat spectrum over a wide range of frequencies.
           Their striking spectra (i.e. spectral index $> -0.5$) clearly distinguishable
   from ordinary radio pulsars (with a spectral index $\sim -1.6$; Camilo et al. 2007a\&b).
   
    \item At their peak, both magnetars are very luminous in radio with 
           luminosity at 1.4 GHz $L_{1.4}\ge 100 {\rm mJy}\  {\rm (d/kpc)}^2$, which is larger than the $L_{1.4}\le 2 {\rm mJy}\ {\rm (d/kpc)}^2$ of most any ordinary young pulsar (e.g.
     Camilo et al. 2006).   For XTE  J1810$-$197, its assumed isotropic radio luminosity
     up to 42 GHz is about $2\times 10^{30}$ erg s$^{-1}$ (Camilo et al. 2006).
          
        \item   Both have variable pulse profiles (exhibiting 
        sudden changes in radio pulse shape) and radio flux densities.
         The flux changes at all frequencies.
     At a given frequency there is no stable average pulse profile. Different pulse
      components change in relative intensity and new components sometimes appear.
       Sub-pulses with typical width approximately $<$ 10 ms are observed (Camilo et al. 2007a\&b).
              
         \item For XTE J1810$-$197, the torque was decreasing, at a time
          when the star was returning to quiescence years after the large outburst.
          As the torque decreased, so did the radio flux (Camilo et al. 2007b).
          
         \item In 1E 1547.0$-$5408, in contrast, the torque has been increasing, at a time when 
          the X-ray flux has been gradually decreasing (Camilo et al. 2007a).

         \end{itemize}
   
   In this paper we extend our existing quark star model for AXPs
   to account for the observed behavior of these two transients.
   This paper is structured as follows: Section 2 gives the basic
     elements of the model. Section 3 describes the quiescent phase.
      Section 4, the bursting phase. Section 5, the radio emission. 
   Model predictions are highlighted in Section 6
    before we conclude in Section 7.

\section{Basic components of the model}

The Quark-Nova  is an explosive transition from a neutron star (NS) to quark star (QS) (Ouyed, Dey, \& Dey 2002;  Ker\"anen\&Ouyed 2003). 
The result is a partial ejection of the NS crust (Ker\"anen, Ouyed, \& Jaikumar  2005) that
 leads to two possible types of debris surrounding the compact remnant (i.e. the QS)
 depending on the QS's birth period. In OLNI, we showed that if the
 QS is born slowly rotating, then the debris formed from the QN ejecta will be in 
co-rotation, which we argue is responsible for SGRs. In OLNII, we showed that for QS born
   with millisecond periods, the debris evolves into a Keplerian
   ring with applications to AXPs.  Furthermore, RRATs are the result of late evolution
    of the Keplerian ring in our model (OLNIII). The interested reader
    is referred to these papers for more details. Below we give an overview
    of the salient features of the model in the case of a Keplerian ring
     before we apply our model to transient AXPs.

\begin{figure*}[t!]
\includegraphics[width=\textwidth]{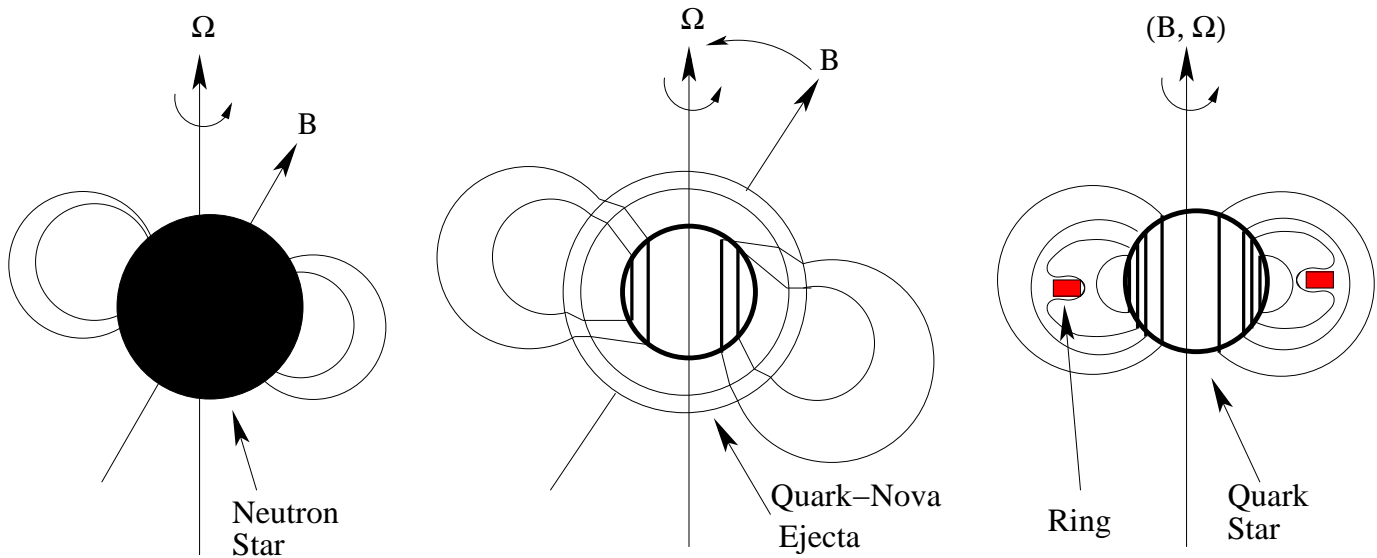}
\caption{
 Illustrated here are the stages involved in the transition from a neutron star (NS) 
  (inclined rotator) to a quark star (QS) (aligned rotator) in the Quark-Nova scenario.
   The collapse of the quark core induces an explosive NS-to-QS  transition 
    ejecting iron-rich degenerate crust material.  The QS enters a superconductive phase
     confining the interior field to vortices, forcing the exterior field to align with  the rotation axis
      (see Ouyed et al. 2004 for more details and Ouyed
       et al. 2006 for the related simulations). The iron-rich degenerate
       ejecta  evolves  into a Keplerian ring  (at about 15-30 km from the star) 
        surrounded, but not penetrated, by the dipole
        field (a co-rotating ejecta is also possible
        depending on the NS period; see OLNI).       
 }
 \label{fig:alignment}
\end{figure*}

\subsection{The quark star: magnetically aligned rotator}

The QN compact remnant is a quark star in the Color-Flavor-Locked (CFL) phase, 
which due to it's rigorously electric neutrality (Rajagopal \& Wilczek 2001) possesses no crust. 
Owing to the superconductivity of the CFL state, the star's interior 
contains a lattice of vortices that confine the magnetic field (Ouyed et al. 2004).  
This interior configuration consequently forces the exterior field 
to align with the rotation axis (Ouyed et al. 2006;  Niebergal et al. 2006);
 this is illustrated in Figure \ref{fig:alignment}.

\begin{figure*}[t!] 
  \label{fig:ringfig}
 \begin{center}
\includegraphics[width=0.8\textwidth]{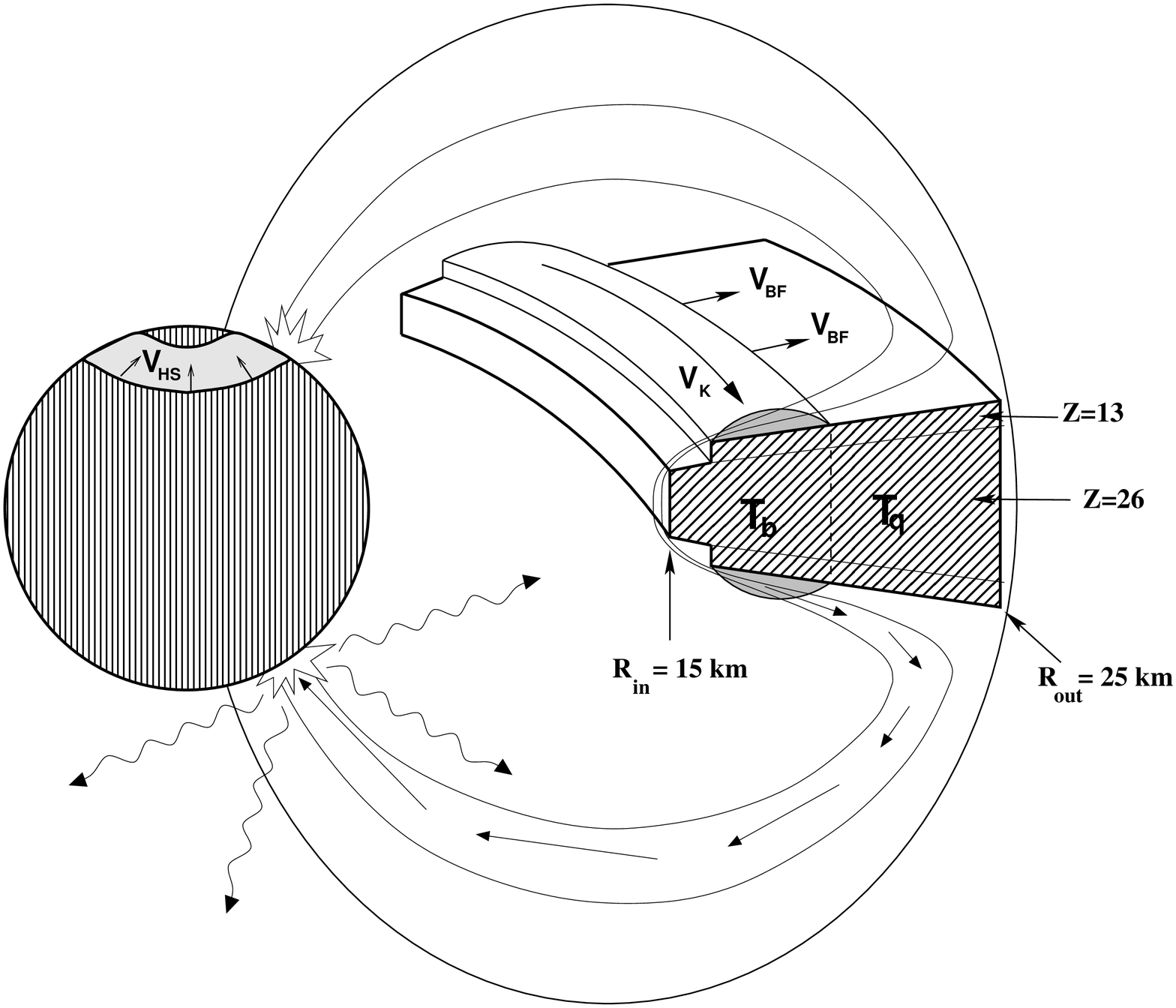}
\includegraphics[width=0.6\textwidth]{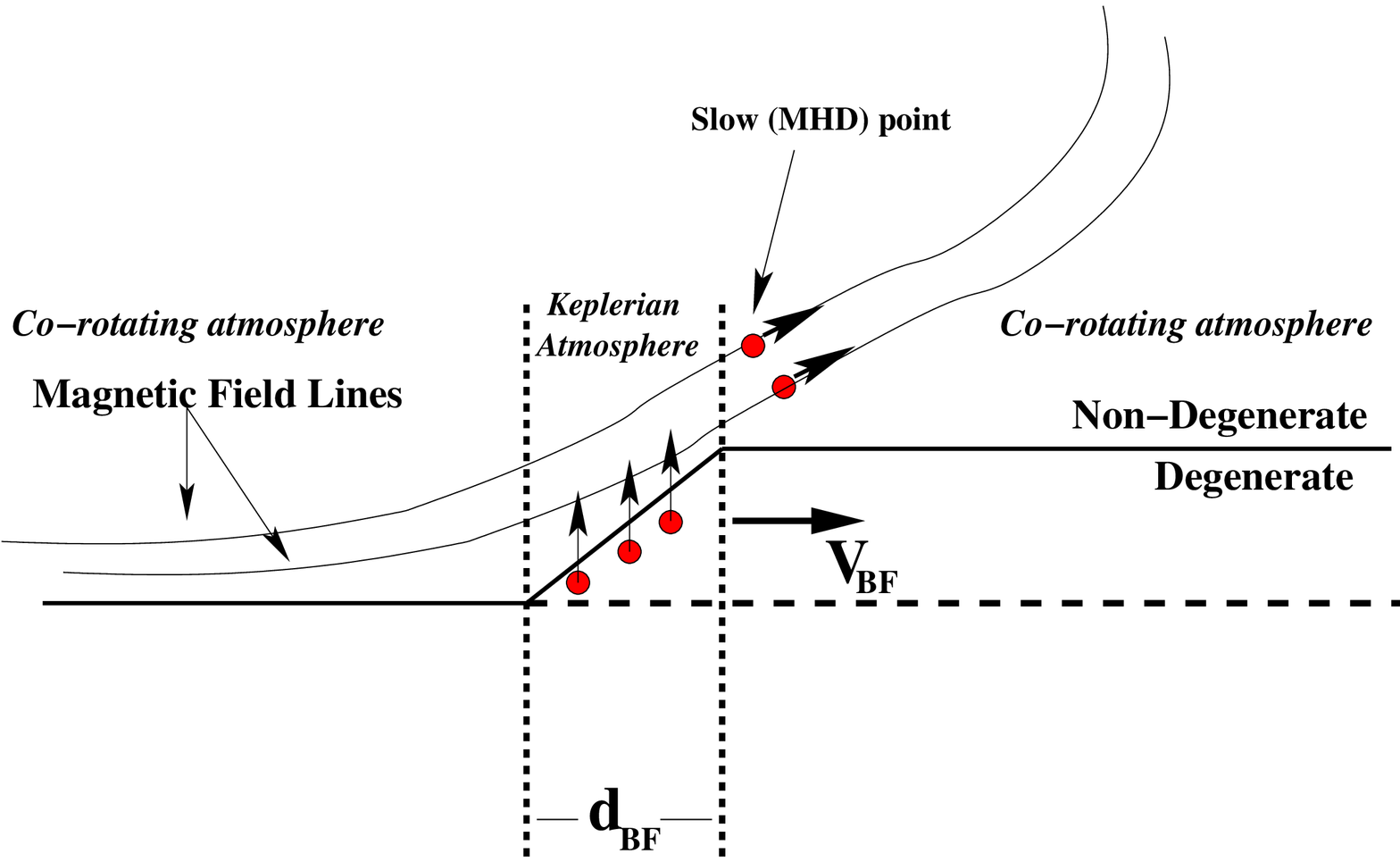}
\caption{
 {\bf The upper panel} 
  illustrates  the ring structure during outburst.
  Fiducial values of the ring inner radius (15 km) and outer radius (25 km)
   are shown; the ring's vertical scale height is a few kilometers (see \S 2.2; the
   figure is not to scale). The ring is surrounded, but not penetrated,  by the dipole field. The  Bohm Front -- ``bump" made of non-degenerate
   Keplerian material -- propagates  outward
    at a speed $V_{\rm BF}$ heating up the system from $T_{\rm q}$ to $T_{\rm b}$
     and inducing accretion onto the star.  The quark star (an aligned rotator) 
      is also shown with the accretion
  hot spot shrinking towards the pole at a speed $V_{\rm HS}$ as a response to the
  outward motion of the accreting Bohm front.  
 {\bf The lower panel} illustrates motion and flow dynamics in the
 vicinity of the Bohm Front (BF).  
The solid line represents the transition to non-degenerate densities
in the ring,  $\rho_{\rm nd}$, while $d_{\rm BF}$ is the radial width
  of the BF. As non-degenerate Keplerian material
   is expelled from the slow point, more Keplerian material
    is supplied from underneath  (see Appendix \ref{app:MHD}). Accretion
    shuts off  once the magnetic field penetrates the non-degenerate
    Keplerian layer re-enforcing co-rotation (behind the BF).
 }
 \end{center}
 
\end{figure*}
    
         \subsection{The Keplerian ring}
     \label{sec:ring}

      Here we are concerned with compact remnants born with millisecond periods.
       As shown in OLNII, the quark star is surrounded by  Keplerian
       debris we refer to as a ring.      
      This Keplerian ring is described in detail in \S 2.1 and Appendix A
       in OLNII.  Briefly,  it is  a high density ($> \sim 10^8$ g cm$^{-3}$) ring 
        rich in iron-group degenerate material (the ejected
       NS crust material) in Keplerian
       rotation around the QS.  
         The ring  is highly 
 conducting, cool, and not threaded by 
  the magnetic field. It is 
 a (possibly amorphous) crystalline solid (like the outer crust of a neutron 
 star).   The ring expands  vertically and radially
        in time to a structure depicted in Figure 2 with the
         ring inner radius at $R_{\rm in}\sim 15$ km, and an outer radius at 
          $R_{\rm out}\sim 25$ km (for these fiducial values the
         total area of the ring is $A_{\rm ring}\sim 3000$ km$^2$).  
         
  The ring thickness
          in the $z$-direction can be shown to be $H_{\rm ring} \sim 2.68\ {\rm km}\rho_{\rm ring, 9}^{1/6} R_{15}^{3/2}$
     where $\rho_{\rm ring,9}$ is the ring's density in units of $10^9$ gm cm$^{-3}$.
  In reality,  on finer scales than depicted in Figure 2, the ring is subject to tidal fracture and is made up
       of many cylinders we refer to as ``walls".  The width
        of each cylinder is set by Keplerian shear resulting  in meter size
         pieces.    
  The mass and  width of a wall  are  given in OLNII and are
      $m_{\rm w}\sim  10^{-10} M_{\odot}$ and 
   $\delta r_{\rm w}\sim 400\ {\rm cm}\ R_{\rm in, 15}^{3/2}$, respectively.   
      
  The magnetic field penetrates a conductor of thickness $\delta r$
  on timescale $\tau_{\rm B}\sim
  (4\pi \sigma/c^2)\times (\delta r)^2$ where $\sigma$ is the conductivity (e.g. \S 4.1 in OLNII).  
 The QS dipole field will  penetrate the  ring via its innermost wall.
  The wall is penetrated   radially because
  $\delta r_{\rm w} << H_{\rm ring}$. This occurs on timescales of
     a few hundred years (see eq.(17) in OLNII).   As the QS magnetic field penetrates
      the innermost wall, magnetic torques (due to induced $B_{\phi}$)
 slow down the wall, so it can accrete (see \S \ref{sec:bursting}).

       \subsection{The ring atmosphere}

        The ring's density decreases with height above the equatorial
         plane.     At any given temperature the ring's density below
      which the ring's matter becomes non-degenerate is found
       by equating the ring temperature to its Fermi temperature; this
       defines the ring's vertical atmosphere. 
  The atmosphere is characterized by its base density, scale height in the $z$-direction
      (vertical to the orbital plane), scale height in the radial direction (i..e.
       on the outer ring edge, $R_{\rm out}$; see appendix B in OLNII), and
        thermal speed, 
 \begin{eqnarray} \label{eq:atmosphere}
 \rho_{\rm atm} &\simeq&  460\ {\rm gm \ cm}^{-3} \ T_{\rm keV}^{3/2}\\\nonumber
 H_{\rm atm}^{z} &\simeq& 20.4\ {\rm cm}\ \frac{T_{\rm keV} R_{\rm  15}^{3/2}}{\mu_{\rm 3.3}}\\\nonumber
 H_{\rm atm, out} &\simeq&  3.7\ {\rm cm}\ \frac{T_{\rm keV} R_{\rm out, 15}^{2}}{\mu_{\rm 3.3}}\\\nonumber
 v_{\rm atm, th} &\simeq& 9.4\times 10^{6}\ {\rm cm\ s}^{-1}\  \frac{T_{\rm keV}^{1/2}}{\mu_{\rm 3.3}^{1/2}}\ ,
 \end{eqnarray} 
  where $T_{\rm keV}$ is the atmosphere temperature in keV, $R_{\rm 15}$ is the
  radial position in units of 15 km, and $\mu_{\rm 3.3}$ is the mean molecular
  weight in units of 3.3 which represents a partially-ionized iron-rich atmosphere
   (see \S 3.2 in OLNII).  The ring atmosphere's scale height in the
    $z$-direction  is to be differentiated from that of the
    ring itself $H_{\rm ring}$  which 
    is of the order of a few kilometers.

      The 
   ring's atmosphere is penetrated by the magnetic field on timescales of $\sim  12\ {\rm days}\ 
   T_{\rm keV, 0.1}^{5/2} R_{\rm in, 15 km}^{3}\mu_{\rm atm., 3.3}^{1/2}$. This can be derived from
     $\tau_{\rm B}$ (as defined \S 2.2) using $\delta r = H_{\rm atm}^{z}$
     and $\sigma\propto 1/c_{\rm s}$ where $c_{\rm s} \propto \sqrt{\mu/T}$
      is the atmosphere's sound speed.
     The ring's temperature is in units
     of 0.1 keV representative of its equilibrium temperature during the quiescent phase;
      see eq(16) in OLNII.
   Since $B^2/8\pi >> \rho_{\rm atm.} V_{\rm K}^2$,
      the atmosphere is forced to co-rotate with the field inhibiting
       accretion onto the star during the quiescent phase.
        On  timescales  of a hundred years, outbursts are triggered by wall accretion.
  These move degenerate ring material into the atmosphere (see \S 4 below) faster than the magnetic field
 lines penetrate vertically  into the
degenerate ring. The ring's penetration  timescale is of the order of tens of millions of years because
 of the $(\delta r)^2$ dependence, so that the ring is not penetrated from above.
       
      In our model,  as we describe in more detail in \S 6.1 
       in this paper, transient AXPs do not accrete during their
       quiescent phase while normal AXPs do accrete continuously from the ring's outer edge,
        where $B^2/8\pi < \rho_{\rm atm.} V_{\rm K}^2$.
         For now we concern ourselves with transient AXPs whose quiescent
          phase is dominated by emission from vortex expulsion as described next.

 \section{The quiescent phase in our model}

 There are two critical radii in our model during the quiescent
      phase, the inner radius $R_{\rm in}$, and the
 outer ring radius $R_{\rm out}$.
  In most cases these radii will be expressed in units of 15 km or 25 km thus
   assigned a subscript 15 or 25. The other 2 parameters
    related to the geometry of the ring are the ring's solid
     angle divided by $4 \pi$ at $R_{\rm in}$ and $R_{\rm out}$ namely,
       $f_{\rm in, \Omega}= H_{\rm ring, in}/R_{\rm in}$ and 
    $f_{\rm out, \Omega}= H_{\rm ring, out}/R_{\rm out}$. 
     General relativistic (GR) effects are included
       in the factors $f_{\rm GR, QS} = \sqrt{1- R_{\rm Sch.}/R_{\rm QS}}$, 
     $f_{\rm GR, in} = \sqrt{1- R_{\rm Sch.}/R_{\rm in}}$, 
      $f_{\rm GR, out} = \sqrt{1- R_{\rm Sch.}/R_{\rm out}}$ while 
      $f_{\rm GR, ring} = 0.5 (f_{\rm GR, in} + f_{\rm GR, out})$ with $R_{\rm Sch.}$
       being the star's Schwarszhild radius.  Unless otherwise specified, quantities
        such as luminosity, temperature, and area are local values.  Values
         at infinity are obtained by using the relevant GR factors.
       Finally, the ring area which includes top, bottom and inner surfaces
        is $A_{\rm out} = 2\pi (R_{\rm out}^2- R_{\rm in}^2) + 4\pi R_{\rm in} H_{\rm ring, in}
      = 2\pi R_{\rm out}^2 Y_{\rm out}$ where  $Y_{\rm out}  =  1 - (R_{\rm in}/R_{\rm out})^2 + 2 (R_{\rm in}/R_{\rm out})^2 f_{\rm in, \Omega} $.

\subsection{The 2 blackbodies in quiescence}
\label{sec:2bbq}

As discussed in OLNI and OLNII, during the quiescent phase we  have 2 blackbodies,
 one from the emission due to magnetic reconnection following
  vortex expulsion (the emission occurs just outside the star's surface).
   The resulting luminosity is  (see \S 5 in OLNI)
   \begin{equation}
   L_{\rm vortex}\sim 2\times 10^{34} \ {\rm erg\ s}^{-1}\eta_{\rm X,0.1}
    \dot{P}_{-11}^2\ ,
    \end{equation}
     with a corresponding temperature, 
\begin{equation}
T_{\rm vortex} \simeq 0.2\ {\rm keV}\ \eta_{\rm X,0.1}^{1/4}  \frac{\dot{P}_{-11}^{1/2}}{R_{\rm QS, 10}^{1/2}}\ , 
\end{equation}
where the period derivative is in units of $10^{-11}$ s s$^{-1}$,  the
star's radius in units of 10 km, 
 and $\eta_{\rm X}$ is the efficiency parameter inherent in the
  conversion from magnetic energy to radiation in units of 0.1.

The  second BB results from reprocessing by the ring of the first BB's X-ray emission,
 $f_{\rm GR, QS}^2 f_{\rm in, \Omega} L_{\rm vortex}= f_{\rm GR, ring}^2  A_{\rm ring}\sigma T^4$.
  The ring-atmosphere system's temperature during quiescence is then,
   since $f_{\rm GR, QS}\simeq f_{\rm GR, ring}$,  
 \begin{equation}
 \label{eq:Tquite}
  T_{\rm q} \sim 0.1\ {\rm keV} \eta_{\rm X, 0.1}^{1/4} \frac{\dot{P}_{-11}^{1/2} }{R_{\rm out, 25}^{3/8} Y_{\rm out}^{1/4}} \ ,
  \label{eq:Tq}
  \end{equation}
  where the subscript  ``q" stands for quiescent in contrast to the values
   during the bursting phase denoted by subscript ``b".  
        These two blackbodies are generic emission components
        to reprocess the emission from the central object,
irrespective of its ultimate origin. As such, they
should also be expected in normal AXPs/SGRs, where the
continuous emission is dominated by the constant accretion
from the ring edge rather than by vortex annihilation.  Since $\dot{P}\propto t^{-2/3}$
    in our model (Niebergal et al. 2006), both BB temperatures during the quiescent phase
     evolve in time as $t^{-1/3}$.

\begin{figure*}[t!]
  \label{fig:compare}
\centering
\includegraphics[width=\textwidth,height=0.25\textheight]{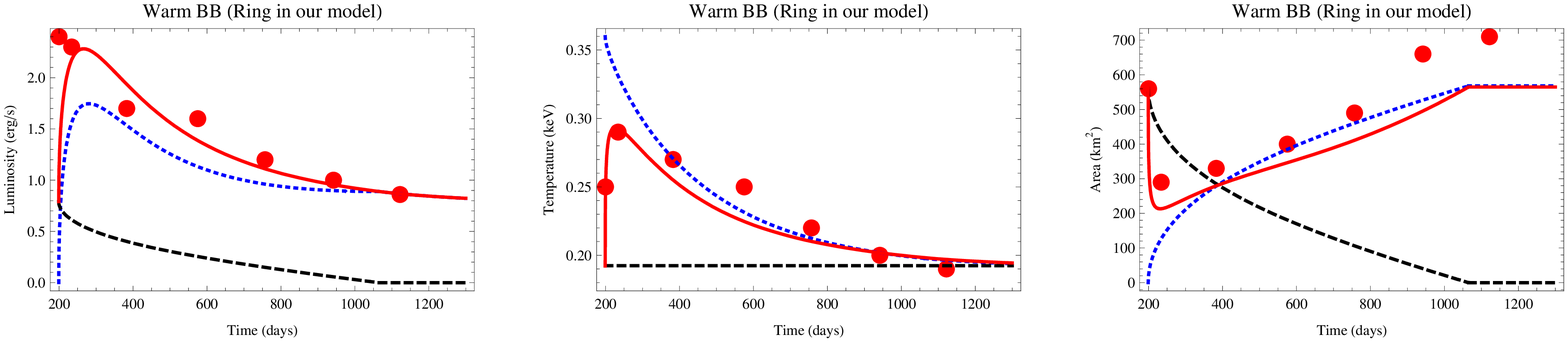}
\includegraphics[width=\textwidth,height=0.25\textheight]{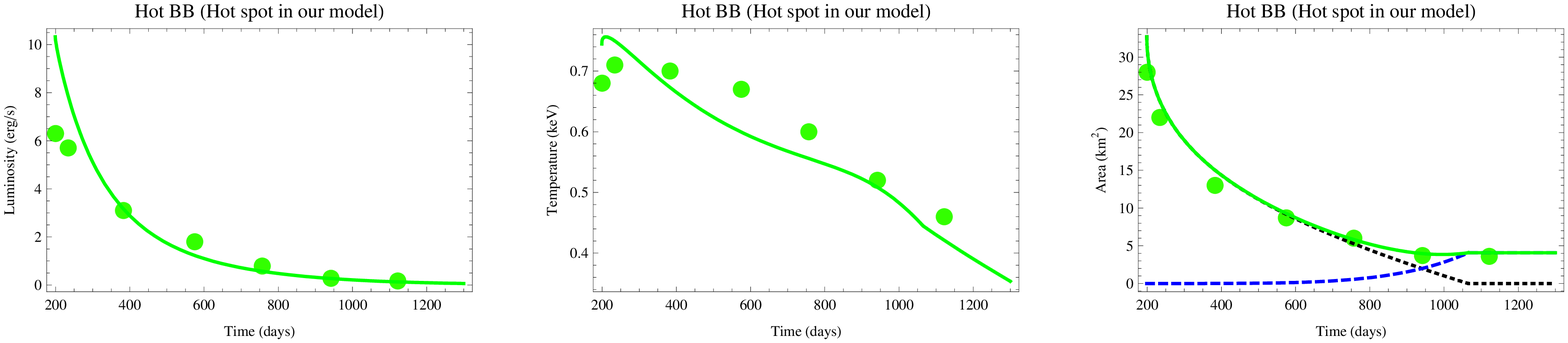}
\caption{{\bf Upper panels}:  Model luminosity ($L/10^{34}$), temperature and area (curves) from
 one side of the ring 
 as observed at infinity compared to observations (dots) of the warm 
component of XTE J1810$-$197.   The dotted line
 is the contribution inward of the BF,  the dashed is from outward
  of the BF, and the solid is the combined contribution (see text); {\bf Lower panels}:
  Model luminosity ($L/10^{34}$), temperature and area (curves) from
 one pole  
 as observed at infinity compared to observations (dots) of the hot  
component of XTE J1810$-$197.   In the right-most  panel
 the dotted line traces the shrinkage of the HS as the BF moves outward, 
  the dashed line shows the late contribution from accretion induced by heating
   of the back side of the ring, and the solid is the combined contribution.} 
\end{figure*}

     The resulting temperature ratio in our model, using a fiducial
      value $Y_{\rm out}\sim 0.64$, is
 \begin{equation}
 \frac{T_{\rm vortex}}{T_{\rm q}} 
 \sim 1.7 \frac{R_{\rm out, 25}^{3/8}}{R_{\rm QS, 10}^{1/2}}\ .
 \end{equation}

   Interestingly a similar correlation between the hot and cool BB  for AXPs
      and SGRs  has been observed (Nakagawa et al. 2009;
see their figure 4), with $T_{\rm BB, H}/T_{\rm BB, C}\sim 2.5$.  This
  we suggest as  evidence for reprocessing in the system. 
  The list of objects
studied by Nagakawa et al. (2009) does not include transient AXPs in quiescence.
   These seem to be  modeled by a single thermal component (e.g. Gotthelf \& Halpern
        2005; Perna \& Gotthelf 2008; Bernardini et al. 2009), although because
    of the low fluxes, presence of the cooler component cannot be ruled out.

\section{Bursting phase in our model}\label{sec:bursting}

\subsection{Consequences of wall penetration and accretion}

The ring remains quiescent until magnetic penetration of the inner
  edge of the ring (wall).   The  magnetic field penetration 
       and subsequent accretion events are very sporadic - they
 last for hours  and occur roughly once every  hundred years (see  section 4 in OLNII). 
  The consequences of wall accretion are as follows:

 \begin{itemize}     
    
    \item  \underline{{\bf Change\ in ring's\ mean\ molecular\ weight}}:
     
  The energy released by the wall accretion is sufficient to dissociate
  a significant mass of iron nuclei in the ring into light nuclei (nuclei
   with $Z\sim13$).   The state of the ring depends on the Coulomb parameter which gives a solidification
temperature of $T_{\rm s} \simeq 9.5 \rho_8^{1/3}$ keV (see \S 3.2 in Ouyed \& Leahy 2009).
 The ring mean density (eqn A.8 in OLNII )
is $\sim  10^7/t^{1/2} \sim 3\times  10^5$ g cm$^{-3}$, with age $t$ in yrs. 
This gives $T_{\rm s} \simeq 1.4$ keV. However
during the  wall accretion event, the ring temperature rises to $\sim 4$-5 keV (see eqn C.2
in  OLNII). This completely melts the ring, which allows light
elements to rise buoyantly. Then the ring re-solidifies on a few hour timescale (eqn
22 of OLNII).  This creates a two layer system as depicted in Figure 2.
             This process reduces the
    molecular weight of the atmosphere from $\mu_{\rm q}\sim 3.3$ to $\mu_{\rm b}\sim 2.1$-$2.5$
     (we adopt an average $\mu_{\rm b}\sim 2.3$).  
   
   As estimated in \S 4.4 in OLNII, for a typical
    wall mass $m_{\rm w}\sim 10^{-10}M_{\odot}$, the number of dissociations 
     following irradiation from wall
      accretion is 
      $N_{13} \sim 10^{46} \zeta_{\rm w, 0.001} \eta_{0.1} m_{\rm w, -10}$
       where $\zeta_{\rm w, 0.001}$ is the dissociation efficiency
        in units of 0.001, and $\eta_{0.1}$ is the
       wall accretion efficiency in units of 0.1. 
       The depth of the $Z \sim 13$ layer can be estimated to be $H_{\rm 13}\sim
              10^{-3} H_{\rm ring}$; that is,  of the order of a few meters. 
       Subsequent depletion of the $Z \sim 13$ nuclei by accretion   leads
    to a return to an iron-rich atmosphere  (i.e. $\mu_{\rm q}\sim 3.3$) as shown
     in OLNII (see also Appendix   \ref{sec:tau13}   in this paper).

 \item \underline{{\bf  The\ Bohm\ front} (BF)}:  
 
 The sudden reheating
  of the inner ring region triggers   heat propagation  outward
  along the ring's atmosphere.  Heat propagation normal to the magnetic
   field can occur by either classical or Bohm diffusion.

The classical diffusion coefficient is 
    $D_{\perp} = 2n\eta_{\perp} k_{\rm B}T c^2/B^2$ where $\eta_{\perp}= 1.44\times 10^-8 Z T^{-3/2}\ln(\Lambda)$ is the transverse Spitzer resistivity with $\ln(\Lambda)$ the Coulomb logarithm
 (e.g. eq. (5.71) in Chen 1984 for S. I. units).  
 The Bohm diffusion coefficient is given by $ck_{\rm B}T/(16eB)$ (see eq(5.111) in Chen 1984  for S. I. units).
The ratio of classical to Bohm coefficients for $B=10^{13}$ G, $T=0.2$ keV and density
 of $10$ g cm$^{-3}$ is of the order of $10^{-8}$ implying 
  that Bohm diffusion dominates over classical diffusion for magnetic
 field strength and temperatures involved here. 
   Thus the heat diffuses outward
  according to Bohm diffusion, introducing 
  a critical radius in our model (during the bursting phase), $R_{\rm BF}=
   R_{\rm in}+ \Delta r$, where $\Delta r$
  is the distance that the Bohm Front (BF) has travelled from $R_{\rm in}$ (see
   Appendix A for details).

   \item  \underline{{\bf Atmosphere\ feeding}}: 
   
   The boundary between the non-degenerate atmosphere and degenerate
       ring material, at the BF,  moves downwards into higher density layers since
         the BF heats up the interface to higher temperatures (recall that $\rho_{\rm atm}\sim
           T_{\rm keV}^{3/2}$; see eq.(\ref{eq:atmosphere})).
         Thus newly  non-degenerate Keplerian ring material is
         unveiled between the co-rotating atmosphere and the interface  (Fig. 2). 
          We thus have two mechanisms that feed 
            the  atmosphere with new non-degenerate
           material. These two contributions are 
             seen when estimating the surface density
              of the atmosphere $\Sigma \sim \rho_{\rm atm} H_{\rm atm}\propto T^{5/2}/\mu$;
               the lower $\mu$ from the buoyancy and the  higher $T$ from the
               heat front will increase the mass up by a factor of $\sim 10$. 
            
   \item  \underline{{\bf Atmosphere\ ejection\ and\ accretion}}:

The huge shear between the co-rotating, magnetized atmosphere and the
 underlying non-degenerate keplerian atmosphere leads to Kelvin-Helmholtz
 instability that helps load the keplerian material onto
 the magnetic field lines attached to the heat front\footnote{During quiescence the
  shear is between the solid degenerate Keplerian ring and the co-rotating atmosphere thus
   not prone to the instability. In  contrast, during
   burst the shear is between one fluid (the  new Keplerian atmosphere)
    and another  fluid (the pre-existing co-rotating 
   atmosphere).}. With simple
 angular momentum  arguments one can show that
 any keplerian particle attached to a co-rotating field line
 will slide along the field line moving radially outward. 
  The magnetic field acts as ramp for the plasma
 particles to get rid of their excess angular momentum.
 This  is the slingshot effect in magneto-hydrodynamic (MHD) jets (discussed in Appendix \ref{app:MHD}).
  In the quiescent state there is a huge shear between the corotating 
 field lines (containing the non-degenerate atmosphere) and the Keplerian 
 degenerate ring material. However since the field lines do not
penetrate the degenerate ring no MHD ejection (and thus no accretion) is feasible. 
         
           In summary, the K-H instability would load mass
 onto the field line attached to the Bohm front and get
 ejected centrifugally  along the field lines as explained in  Appendix \ref{app:MHD}
           (see also below).          
 The wind is then channeled onto the star with an accretion rate, $\dot{m}_{\rm acc.}$,  given by equation(\ref{eq:mdotacc}) creating a hopt spot (HS) on the surface of the star.

 \end{itemize}

   \subsection{The 2 blackbodies during burst}
 \label{sec:2bbb}    
 
 Only magnetic field lines that
 are in the path of the heat front get loaded and accrete onto
 the star.  The main consequence,
  as we show in details here,  is that the HS on the star moves closer
to the pole thus decreasing in area.

      The temperature of the inner ring is obtained by equating heating from BF induced
       accretion onto
    the quark star 
    with blackbody cooling by the inner ring surface ($R < R_{\rm BF}$); $f_{\rm GR, QS}^2 f_{\rm BF,  \Omega} L_{\rm acc.}= f_{\rm GR, ring}^2 A_{\rm BF}\sigma T_{\rm  b}^4$ where $L_{\rm acc.}=\eta \dot{m}_{\rm acc.} c^2$
     is the accretion luminosity, $\eta$ the accretion efficiency, $f_{\rm BF, \Omega}=H_{\rm ring, BF}/R_{\rm BF}$, and the area created by the Bohm front is $A_{\rm BF} = 2\pi (R_{\rm BF}^2- R_{\rm in}^2) + 4\pi R_{\rm in} H_{\rm ring, in} = 2\pi R_{\rm BF}^2 Y_{\rm BF}$.  Here  $Y_{\rm BF}  =  1 - (R_{\rm in}/R_{\rm BF})^2 + 2 (R_{\rm in}/R_{\rm BF})^2 f_{\rm in, \Omega}$.  
     The resulting temperature is,
\begin{equation}
\label{eq:Tequi}
T_{\rm b} \simeq 0.21\ {\rm keV} \ \eta_{0.1}\zeta_{\rm sp, 0.01} \frac{R_{BF, 15}}{\mu^{3/2} Y_{\rm BF}}\ .
\end{equation}
 Here, $\zeta_{\rm sp, 0.01}$  in units of 0.01, is a factor related to mass-loading
  at the slow MHD point as determined in Appendix B.
The approximation above is valid until late times when
 accretion ceases, so that $T_{\rm b}$
  is no longer given by the accretion-cooling balance but rather settles
   to the quiescent temperature $T_{\rm q}$.  The outer ring's temperature is 
given by the quiescent temperature, eq.(\ref{eq:Tq}), until the heat front reaches it.
  
  The accretion rate is then obtained by combining eq.(\ref{eq:mdotacc})
   with equation above to get
   \begin{equation}
   \dot{m}_{\rm acc.}\sim 1.6\times 10^{15}\ {\rm g\ s}^{-1}\ \eta_{0.1}^3\zeta_{\rm sp,0.01}^4 \frac{R_{\rm BF, 15}^{11/2} f_{\rm K}(T_{\rm b})}{\mu^6 Y_{\rm BF}^3}\ .
   \end{equation}
   The function $f_{\rm K}(T_{\rm b})$  is described in Appendix \ref{app:MHD} and acts as a ``valve" that 
          shuts-off accretion after the heating front has passed.
   There is a fixed amount of Keplerian material at any radius  that
          can be fed into an MHD wind and thus accreted onto the star.
           This is because of re-enforced co-rotation once the magnetic field penetrates
            the non-degenerate Keplerian layer effectively shutting off accretion (see
             Figure 2). 
            
            The natural connection between the
      BF outward expansion and the  resulting HS on the star  is  illustrated in Figure
       2.
   The corresponding HS luminosity is 
   \begin{equation}
   \label{eq:hslum}
   L_{\rm acc.}\sim 1.4\times 10^{35}\ {\rm erg\ s}^{-1}\ \eta_{0.1}^4\zeta_{\rm sp,0.01}^4 \frac{R_{\rm BF, 15}^{11/2}f_{\rm K}(T_{\rm b})}{\mu^6 Y_{\rm BF}^3}\ ,
   \end{equation}   
   with a HS temperature 
    \begin{equation}
    T_{\rm HS}\sim 0.58 \ {\rm keV}\ \eta_{0.1}\zeta_{\rm sp, 0.01} \frac{R_{\rm BF, 15}^{11/8}f_{\rm K}(T_{\rm b})^{1/4}}{\mu^{3/2} Y_{\rm BF}^{3/4} A_{\rm HS, 120}^{1/4}}\ ,
    \end{equation}
     where the HS area, $A_{\rm HS}$  is in units of 120 km$^{2}$, is  derived in the Appendix (see eq. \ref{eq:hsarea}).

     In summary, during the burst phase,  the ring-atmosphere system consists of 2 BBs; the
 inner warmer part  increasing in area at the Bohm diffusion rate 
  and the outer cooler part  decreasing in area. 
  The  hot BB in our model is provided by  the HS which
  decreases in area as the BF propagates outwards.

      \subsection{Temperature ratio during burst}
  
  The temperature ratio in the bursting phase case is then:
 \begin{equation}
 \frac{T_{\rm HS}}{T_{\rm b}} = (\frac{A_{\rm BF}}{f_{\rm BF,  \Omega}A_{\rm HS}})^{1/4}
 \sim 2.8 \left(\frac{R_{\rm out, 25}}{R_{\rm QS, 10}}\right)^{1/2}\ ,
 \end{equation}
  where we approximated $A_{\rm BF}\sim A_{\rm ring}/2$, $f_{\rm BF,  \Omega}\sim f_{\rm out,  \Omega}$, and $A_{\rm HS}\sim 0.1 A_{\rm QS}$ to get the numerical value.
  Our model
   predicts the temperature ratio during burst for a given object to be higher by a factor of $\sim 1.5$
    than  during quiescence.  We expect a scatter in comparing different objects
    mainly  caused by variations in $R_{\rm out}$.   This is consistent  with Figure 4 of Nakagawa et al. (2009).

  \subsection{Application to XTE J1810$-$197}

    The upper panels in Figure 3 compares the time evolution of the ring's
    luminosity, temperature and area in our model 
     to the observed warm BB in XTE J1810$-$197. In our
      model the luminosity from the inner  part of the ring is $L_{\rm BF}=A_{\rm BF}\sigma
       T_{\rm b}^4$ while the outer part gives $L_{\rm out}=A_{\rm out}\sigma
       T_{\rm q}^4$ with $A_{\rm out}= 2\pi (R_{\rm out}^2- R_{\rm BF}^2)$.
        The total luminosity from the ring during burst is $L_{\rm ring, b}= L_{\rm BF}+L_{\rm out}$
         while the corresponding effective area and temperature were derived by weighting 
          over  luminosity, $A_{\rm ring, eff.} = (A_{\rm BF} L_{\rm BF} + A_{\rm out} L_{\rm out})/L_{\rm ring, b}$ and $T_{\rm ring, eff.} = (T_{\rm b} L_{\rm BF} + T_{\rm q} L_{\rm out})/L_{\rm ring, b}$.  
          
          The lower panels in Figure 3 compare the time evolution of the HS's
    luminosity, temperature and area in our model 
     to the observed hot BB in XTE J1810$-$197. In this case,  only one  component 
      comes into play, the accretion luminosity $L_{\rm acc.}$. 
      However,  as the BF gets closer to $R_{\rm out}$ the back side of the ring
       is heated resulting in additional accretion by the same mechanism (i.e.
       eating into Keplerian material in the radial direction this time) and additional area on the polar cap 
        defined by the field lines connecting the back side of the ring to the star (see
        Figure 2).  The resulting area is 
       $A_{\rm HS}+ A_{\rm edge}$  and temperature $ (L_{\rm acc.}/(\sigma (A_{\rm HS}+ A_{\rm edge})))^{1/4}$.  
       
       The fits to XTE J1810$-$197 data were obtained for the following set of parameters:
       \begin{eqnarray}
       R_{\rm in} &=& 23.5\ {\rm km}\\\nonumber
       R_{\rm out} &=& 26.5\ {\rm km} \  ,
       \end{eqnarray}
       and by slightly adjusting the mass-load
         at the slow MHD point so that
         $\zeta_{\rm sp}=0.02$ (see discussion following eq.(\ref{eq:b1})). 
         The other parameters were kept to their fiducial values including
       the star's parameters  $R_{\rm QS}=10$
       km, $M_{\rm QS}=1.4 M_{\odot}$. The star's magnetic field is given
        by $B_{\rm QS}=\sqrt{3 \kappa P \dot{P}} \sim 3.3\times 10^{14}$ G (with $\kappa= 8.8\times 10^{38}$ G$^2$ s$^{-1}$
         as given in eq.(10) in OLNIII).         
   
    Assuming that XTE J1810$-$197 has experienced a few bursting events,
    using equation (A.7) in OLNII, the ring would have spread to no more than
  $(\Delta r)_{\rm t}\sim 8\ {\rm km}$ if the system's 
   temperature during quiescence remained on average $\sim  0.1$ keV. 
   This is consistent with ($R_{\rm out}-R_{\rm in}$)
     found from fits to the XTE J1810$-$197 data thus providing a self-consistency check on our model.
     This also confirms our overall findings in previous work (OLNI, OLNII and OLNIII)
    that the ring should be a  few kilometers in width after a few hundred years.

       \subsection{Application to 1E 1547.0$-$5408} 
       
       In its bursting phase, this
         source was fitted with a $T\sim 0.5$ keV hot  BB   with
         $L_{\rm BB}\sim 1.3\times 10^{35} (d/9\ {\rm kpc})^2)$ erg s$^{-1}$
          and   a corresponding area  decreasing from 180 km$^2$  in June 2007
          to 96 km$^2$ in August 2007 (see  Halpern et al. 2008).
            During the quiescent phase, $L_{\rm BB}\sim  10^{34} (d/9\ {\rm kpc})^2)$ erg s$^{-1}$, 
            $T_{\rm BB}\sim 0.4$ keV and $A_{\rm BB}\sim 36$ km$^2$ (see Table 1 in
            Halpern et al. 2008).  
          
          This source has not been as well sampled as was XTE J1810$-$197.
           Nevertheless, in our model,  high BB temperatures during
            the quiescent phase are suggestive of a  more compact ring
             which is closer to the star (see eq.(\ref{eq:Tquite})).
              In our model, such a small and compact ring could have been
               a consequence of a smaller amount of crust material ejected during the
                QN (see eq.(2) in OLNII).
                         For example, $R_{\rm out}= 15$ km and  $R_{\rm in}=13$ km,
                          which implies $Y_{\rm out}\sim Y_{\rm BF}\sim 0.117$, 
                          inserted in eq.(\ref{eq:Tquite}) gives $T_{\rm q}\sim 0.35$ keV;
                           we take $\dot{P}_{-11}\sim 2$ s s$^{-1}$ for this source.
                           The ring's burst epoch temperature is then $\sim 0.48$ keV
                            with a corresponding peak luminosity from eq.(\ref{eq:hslum}) of
                            $L_{\rm acc.} \sim 4.8\times 10^{35}$ erg s$^{-1}$ for $\mu_{\rm b}=2.4$.
                             This is close to the $\sim 1.3\times 10^{35} (d/9\ {\rm kpc})^2)$ erg s$^{-1}$ measured in 
            June-July 2007 (the peak   of the outburst was not observed and could have been
           higher than this).  
           The initial area of the HS  is given by equation (\ref{eq:hsarea})
                  and is estimated to be $\sim 80$ km$^2$ using $R_{\rm in}=13$ km and $R_{\rm out}=15$ km.
                   Finally, in the case of 1E 1547.0$-$5408 there seems to be hints
        of an aligned rotator from its small X-ray pulsed fraction and its relatively broad 
         radio pulse. In our model, the QN compact remnant (the QS) is born as
          an aligned rotator due to the vortex confinement of magnetic field (see Figure \ref{fig:alignment}).

\section{The radio emission in our model} 

In this section we develop a scenario for radio emission.
 As discussed in \S 1.2,  XTE J1810$-$197 and 1E 1547.0$-$5408
  show unique characteristics in radio (including delay after X-ray
   outburst, a flat spectrum, and unusual spin-down behavior). 
    In chronological order the sequence of events that  leads
     to radio emission in our model is as follows:

   \begin{itemize}
   
    \item     Magnetic bubbles are generated at the ring during the X-ray outburst (see \S 5.1).

      \item  The propagation time of these bubbles from the site of production (the ring) 
      to the l.c. defines the delay between radio and X-ray outbursts (see \S 5.2).

       \item  The unusual spin-down behavior is induced
       by torques on the magnetosphere from the co-rotating bubbles, during their outward
        propagation (see \S 5.5).

      \item  Relativistic collisionless reconnection at the l.c. destroys the bubbles.

  \item  Magnetic energy released by the bubbles is
eventually radiated by particles accelerated by
reconnection at the l.c., mostly at radio
frequencies (\S 5.3).
We suggest that the flat radio spectrum is naturally
associated to this mechanism (see \S 5.4 below).

        \end{itemize}

\subsection{Magnetic reconnection and bubble generation}

After  penetration, the poloidal magnetic field ($B_{\rm p}$) lines
inside the  wall (inner ring)  are dragged by the Keplerian
 shear generating a toroidal magnetic field, $B_{\phi}$.
   Continuous reconnection events during the winding
    of the field lines lead to X-point generation and emergence of closed magnetic bubbles; 
more concisely these magnetically confined plasma bubbles (plasmoids)
       result from the Keplerian shear.

       Generation of bubbles is a common
 feature of threaded disks as seen in many simulations (e.g.
 Romanova et al. 1998; Yelenina et al. 2006). These simulations show the magnetic bubbles
 to acquire enough speed to escape gravity and 
expand freely outward. Magnetic loops and bubbles 
  are also common in the sun (referred to as plasmoids)
 and are also found to gain enough energy from the reconnection
 events to escape the system and expand freely (e.g. Wagner 1984;
 Tamano 1991). 

One key difference between these cases and our model is the
fact that in our model, during quiescence, the degenerate ring is not
 threaded by the magnetic field.
The bubbles form only during
 the bursting phase once the magnetic field has penetrated
 the inner ring and is sheared.  The B field attached to  
the broken inner pieces of the wall is wound up by the keplerian angular 
 velocity ($\sim 6000$ rad/s) of the piece vs. the corotation angular velocity 
($\sim 1$ rad/s) of the footpoint of the fieldline on the
quark star. This rapid field line winding results in reconnection and magnetic 
 loop (bubble) formation. The B field penetration (and subsequent wall accretion with concurrent
 bubble generation) events are very sporadic
 and unique to our model - they
 last for about an hour and occur once every century (see \S 4 in OLNII).

      A rough estimate of the  number of bubbles  that can be generated is
      \begin{equation}
      N_{\rm Bub.}\sim \frac{V_{\rm w}}{V_{\rm Bub.}}\sim  {8.3\times 10^{6}}{R_{\rm in, 15}^{3/2}}\ ,
      \end{equation}
       where the total reconnection volume is $V_{\rm w}\sim 4\pi R_{\rm in} H_{\rm ring, in}  \delta r_{\rm w}$
        and $V_{\rm Bub.}\sim (4\pi/3) \delta r_{\rm w}^3$. 
       The mass of a given bubble is  thus 
   $m_{\rm Bub.}\sim (4\pi/3)\delta r_{\rm w}^3 \rho_{\rm atm., w}$
    where the plasma confined by the bubbles has  a density at birth 
     given by the atmosphere density 
     $\rho_{\rm atm.,w}$.  The magnetic energy stored in each bubble
      is then $E_{\rm Bub.}\sim (4\pi/3)\delta r_{\rm w}^3 (B_{\phi}^2/8\pi)$
       with $B_{\phi}\sim B_{\rm p, in}$ with  
    $B_{\rm p, in}= B_{\rm QS} (R_{\rm QS}/R_{\rm in})^3$.

 \begin{figure*}[t!]
   \begin{center}
\includegraphics[width=\textwidth]{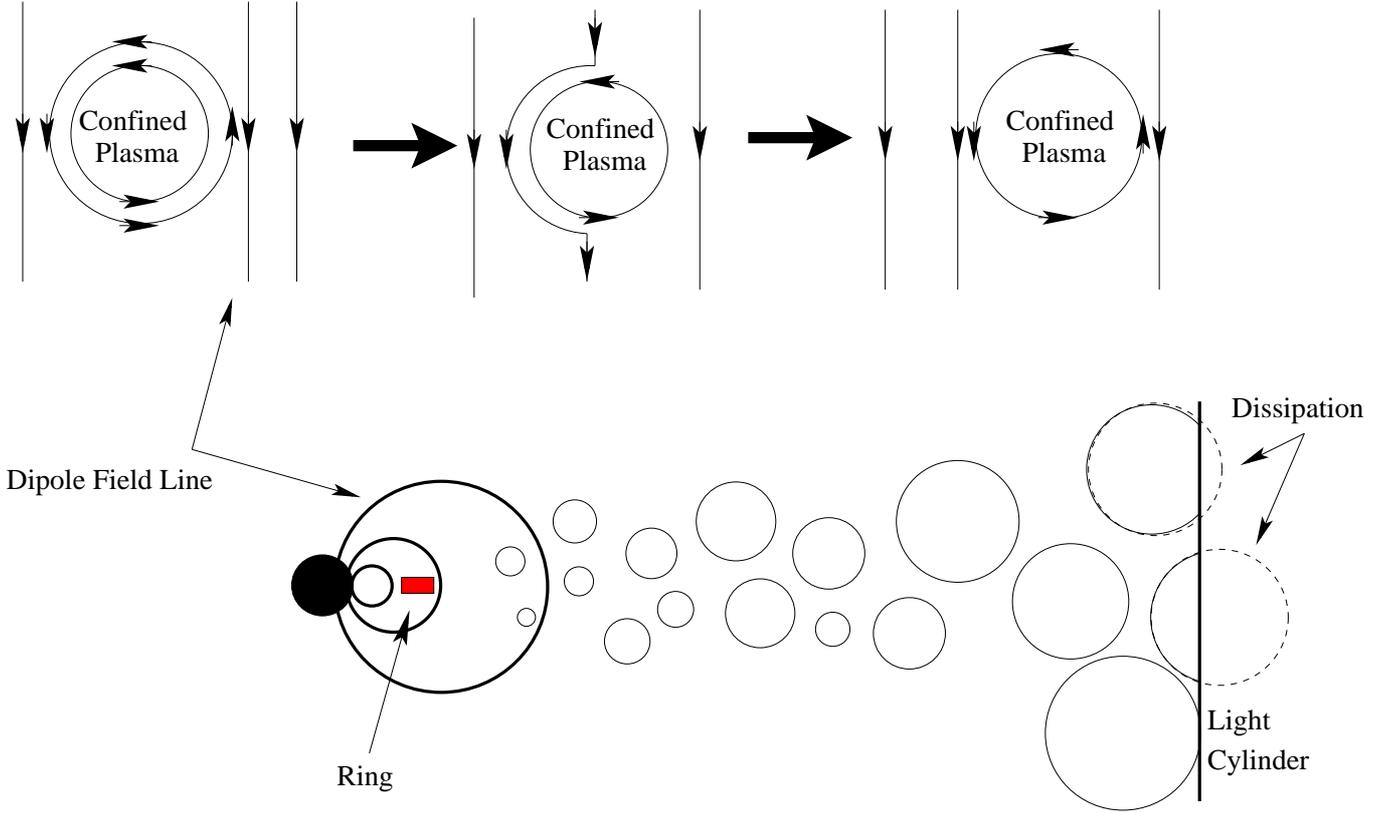}
\caption{Keplerian shear during wall accretion  leads to X-point reconnection which in turn 
causes the generation of closed magnetic loops, or ``bubbles'' (see \S 5.1 in text).  In our model,
these magnetic bubbles are responsible for the radio emission seen in some AXPs.
The top panel illustrates the outward migration of a bubble through the surrounding
magnetic field.  The bubble's magnetic polarity in front as it moves outward
 is opposite to that of the surrounding dipole. This leads to reconnection and the
  motion of the  bubble towards the light cylinder.
The lower diagram shows the overall pattern
of bubble migration (i.e. magnetic buoyancy) and expansion that leads to pile up
      near the light cylinder. The bubbles dissipate as they cross
 the light cylinder leading to the radio emission (see text for details).     
 }
\end{center}
\end{figure*}
           
     \subsection{Delay between the X-ray and radio emission}

     Simulations of bubbles in threaded disks show that  magnetic loops and bubbles propagate outward because of the
 energy they gain from reconnection and because of 
the dipolar B-field gradient from the central  star, similar to the way
solar field disturbances propagate outward above the solar photosphere.
These bubbles  propagate at the sound speed of non-dissipative magnetic field disturbances, 
    ie. at the Alfv\'en speed. 
Illustrated in Figure 4 in this paper, is a rudimentary representation
    of what is seen in these simulations; an  
 outward migration process  through the dipole field involving
  reconnection.  This is a simple
   and only  qualitative model for bubble propagation, which is a 
propagating  geometric disturbance in the magnetic field/plasma. Bubble propagation 
results  in different field lines and different plasma making up the bubble as it 
moves. As seen from figure 4, a magnetic field line in front of the 
bubble is  only temporarily  part of the bubble, and leaves the bubble after
the bubble has propagated past it. The reconnection on
the front side of the bubble results in surface pressure similar to the 
static  $B^2/8\pi$, so the dynamics of reconnection during bubble migration are 
important to how the bubble evolves (both its speed and change in  volume) as 
it migrates.

 The Alfv\'en velocity depends on the
     magnetic field geometry  inside the light cylinder (which
      we take to be nearly dipolar) and on  the ambient density
       which we take to scale as  $\propto  (R_{\rm in}/r)^{\alpha}$.
       If there is a steady wind inside the magnetosphere, then if the wind is spherically symmetric the density we expect would decline as $r^{-2}$, or if it is equatorially confined the density would decline as $r^{-1}$;  that is  $1 \le \alpha \le 2$.  Although the exact distribution is not  at all
       known  (e.g.  Michel 1969; Spitkovsky 2008; 
       see more discussion on this aspect of our model 
       in \ref{sec:limit}) we will adopt    $\alpha =1$ as we expect
        the bubbles to remain along the equator since they are Keplerian at birth. 
   
   The bubbles move outwards at a rate given by  $dr/dt = v_{\rm A}$ where 
    $v_{\rm A} = c (R_0/r)^{3-\alpha/2}$  so that 
         $R_0\sim 45\ {\rm km} \left( B_{\rm QS,14}/T_{\rm atm, keV}^{5/4}\right)^{1/(3 - \alpha/2)}$ 
        is the radius beyond which the Alfv\'en speed becomes sub-relativistic; the star's magnetic field $B_{\rm QS,14}$     is in units of $10^{14}$ G.

      The time it takes the magnetic bubbles to reach the light cylinder (lc)  at $R_{\rm lc}=c/\Omega$ (at which
       point they start dissipating) is found from $\int_{R_0}^{R_{\rm lc}}  dr/v_{\rm A}$,
       \begin{equation}
       t_{\rm prop.}\sim \frac{1}{4-\frac{\alpha}{2}} \left( \frac{R_{\rm 0}}{c}\right) \left( \frac{R_{\rm lc}}{R_{\rm 0}}\right)^{4-\frac{\alpha}{2}}\propto \frac{P^{\frac{7}{2}-\frac{\alpha}{2}}}{\dot{P}^{1/2}}\  ,
       \end{equation}    
       where we made use of the definition of $R_{0}$ and  
         $B_{\rm QS}\propto  \sqrt{P \dot{P}}$ in our model.
      For example for $\alpha =1$,  we get
      \begin{equation}
      t_{\rm prop.} \sim 4.1\ {\rm yrs}\ \frac{P_5^{3}T_{\rm atm.,keV}^{5/4}}{\dot{P}_{-11}^{1/2}} \ ,
      \label{eq:radiodelay}
      \end{equation}
      where the period is given in units of 5 seconds;
       recall that 
         $B_{\rm QS}= 3.6\times 10^{14}\ {\rm G}\ \sqrt{P_5\dot{P}_{-11}}$.
      As they slow down while flowing towards the lc
          the bubbles will pile-up. This implies a delay between the X-ray outburst following wall accretion  and the peak of radio emission in our model.
       Thus, following the peak of the X-ray outburst for XTE J1810$-$197 (with $P\simeq 5.54$ s) and 1E 1547.0$-$5408 (with $P\simeq 2$ s),
          we expect a corresponding radio peak emission to occur  with a delay of 
           $\sim 320$ days and   $\sim 15$  days respectively (assuming $T_{\rm atm.}\sim 0.3$ keV
            and $\dot{P}_{-11}\sim 2$).

\subsection{Duration and luminosity}
         
        The radio duration has two components in our model:
           (i)  The radiative lifetime
         of the electrons, $t_{\rm cool}$;                           
           (ii)  The time difference between when the first bubble arrives and the
 last bubble arrives at the l.c., $t_{\rm radio} = \Delta t_{\rm prop.}$. As we show at the end of this section,
  propagation delays between bubbles is the dominant component.
   Below we focus on case (ii).

         Since all bubbles are produced within a few hours (during the X-ray burst) and at the
 same location, the radio duration is determined by variations in propagations times
 of the bubbles to the l.c..
 The magnetic reconnection events lead to variations in $B$ which
  translate to fluctuations in $v_{\rm A}$ and thus propagation time (consequently arrival
 time at the l.c.) resulting in 
  $ t_{\rm radio} = \Delta t_{\rm prop.} = (t_{\rm prop.,max} - t_{\rm prop.,min}) = t_{\rm prop} \times (\Delta B/B)$ 
 where $(\Delta B/B) = (B_{\rm max} - B_{\rm min})/B$. Or, 
        \begin{equation}
        t_{\rm radio}\simeq  t_{\rm prop.}\times \frac{\Delta B}{B} \ .
        \end{equation} 
   For a homologous expansion  during the outward propagation,
 a given bubble increases in size as $\delta r_{\rm Bub.}\sim \delta r_{\rm w} (r/R_{\rm in})$.
  Combined with flux conservation, $B_{\rm Bub.}\propto \delta r_{\rm Bub.}^{-2}$
   and $V_{\rm Bub.}\propto \delta r_{\rm Bub.}^{3}$,  it yields $E_{\rm Bub.}=  V_{\rm Bub.}B_{\rm Bub.}^2/(8\pi)\propto r^{-1}$.
           The radio luminosity is then given as, for $\alpha =1$, 
     \begin{eqnarray}
     L_{\rm radio}&\simeq&  \eta_{\rm R}  \frac{E_{\rm mag., l.c.}^{\rm Bub.}}{t_{\rm radio}} \\\nonumber
     &\sim&       
     4.4\times 10^{29}\ {\rm erg\ s}^{-1}\times  \eta_{\rm R} \times \frac{B}{\Delta B}\times \\\nonumber 
     && \frac{\dot{P}_{-11}^{3/2} R_{\rm QS,10}^6 \rho_{\rm ring,9}^{1/6} }{P_5^{3}R_{\rm in, 15} T_{\rm atm.,keV}^{5/4}} \  ,     
     \label{eq:Lradio}
     \end{eqnarray}
     where  $E_{\rm mag.,l.c.}^{\rm Bub.} = E_{\rm mag., in}^{\rm Bub.}\times (R_{\rm in}/R_{\rm l.c.})\sim  (\frac{B_{\rm p, in}^2}{8\pi} V_{\rm w})\times (R_{\rm in}/R_{\rm l.c.})$ is the total 
     magnetic energy stored in the bubbles  by the time they reach the light cylinder.
      The ratio $(R_{\rm in}/R_{\rm l.c.})$ is the dilution factor induced by the
       homologous expansion of the bubbles as they propagate from $R_{\rm in}$ to $R_{\rm l.c.}$;
        recall that $B_{\rm in}= B_{\rm QS}(R_{\rm QS}/R_{\rm in})^3$. 
          In the equation above,
       $\eta_{\rm R}$ is the efficiency of conversion of magnetic
       energy to radio emission via reconnection   (e.g. Ouyed et al. 2006) at the l.c..

           For XTE J1810-197, Camilo et al. (2007c) report radio emission
 lasting at least 272 days with initial fading from 5 mJy to 2mJy in the
first few weeks (see their figure 2).  In our model,  $t_{\rm radio} \sim 272$ days gives
 $\Delta B/B\sim 0.85$. Then using $R_{\rm in}\simeq 23.5$ km (from the X-ray
  fits in \S 4.4), 
 $L_{\rm radio}\simeq \eta_{\rm R}\times 3\times 10^{30}$ erg s$^{-1}$. This matches
  the observed radio luminosity of $\sim 2\times 10^{30}$ erg s$^{-1}$ for $\eta_{\rm R}\sim 0.7$
   which is suggestive of a high dissipation/reconnection efficiency of the bubbles at the l.c..

For 1E 1547-5408,   radio emission was observed  (Camilo et al. 2008) 
 from June through August 2007 ($\sim 90$ days). It was not detected Jan 22, 2009  within 18 hours
  of the first reports of renewed X-ray (SGR) bursting activity from it (Camilo et al. 2009).
    In our model,  this is suggestive of 
 $\Delta B/B\sim 6.0$. Then using $R_{\rm in}\simeq 13$ km (from the X-ray
  fits in \S 4.5), 
 $L_{\rm radio}\simeq \eta_{\rm R}\times 2\times 10^{32}$ erg s$^{-1}$. This matches
  the observed radio luminosity of $\sim 2\times 10^{30}$ erg s$^{-1}$ for $\eta_{\rm R}\sim 1\%$.

  The fits above indicate an order of magnitude difference in efficiency, $\eta_{\rm R}$, between the two sources. However, a smaller $\alpha$ (i.e. a slower density decrease inside the l.c.)
   for 1E 1547-5408 would lead to a longer propagation time and lower radio luminosity,
    implying a higher reconnection efficiency.  Alternatively, a more compact quark star
     for 1E 1547-5408 would also imply a higher reconnection efficiency.

         As a given bubble reaches the l.c. it is sheared and spread-out around
          the equator of the l.c..  The resulting bubble's magnetic field  is
           given by $B_{\rm l.c.}^{Bub.}(\delta r_{\rm l.c.}^{Bub.})^2= B_{\rm sh.}^{Bub.}(\delta r_{\rm l.c.}^{Bub.})\times 2\pi R_{\rm l.c.}$. This yields
           \begin{equation}
           B_{\rm sh.}^{Bub.} \sim 60\ {\rm G} \frac{\dot{P}_{-11}^{1/2} R_{\rm QS,10}^3}{P_5^{3/2}R_{\rm in,15}^{1/2}}\ .
           \end{equation}
           After shearing, reconnection occurs accelerating electrons (see \S \ref{sec:reco}).
            The electrons emit synchrotron radiation with a 
          critical  frequency of $\nu_{\rm c}= 0.42 {\rm GHz} \gamma_{10}^2 B_{\rm sh., G}^{\rm Bub.}$ where $\gamma_{10}$ is the  electron  Lorentz factor in units
          of  10  following reconnection and   the sheared 
          bubble magnetic field is in units of Gauss.  We get
           \begin{equation}
           \nu_{\rm c} \sim 25 \ {\rm GHz}\ \gamma_{10}^2 \frac{\dot{P}_{-11}^{1/2} R_{\rm QS,10}^3}{P_5^{3/2}R_{\rm in,15}^{1/2}}\ .
           \end{equation}
              For XTE J1810$-$159 the above gives $ \nu_{\rm c} \sim 24 \ {\rm GHz}$
               while it is $ \nu_{\rm c} \sim 150 \  {\rm GHz}$ for 1E 1547-5408.            
              Observations at higher frequencies than currently available 
               could constrain our model.

              The above calculations neglect   the radiative lifetime
         of the electrons.  However in cases where 
          the bubbles reach the l.c.  simultaneously (i.e. with negligibly short time
delays),   the radio duration will be dominated by   the radiative cooling lifetime of the electrons,  
          $t_{\rm cool} \sim 0.25\ {\rm day}/ (\gamma_{10} B_{\rm sh., 60}^2)$ (Lang 1999). 
                For our estimated values of $\gamma$ and $B_{\rm sh.}$, the propagation delay is
much longer than the electron radiative lifetime. E.g. for 1E 1547-5408, the 
radiative lifetime only becomes important, with $\gamma=10$, for $B_{\rm sh.} <3$ G,
which would also result in the critical synchrotron $\nu_{\rm c}$ below the observed radio
frequencies.

      \subsection{The flat spectrum}
      \label{sec:reco}
    
    As they cross the lc,  the bubbles will 
     dissipate by braking up into smaller bubbles while driving
      complex, non-linear reconnection events. 
     Particle in Cell simulations of  X-point reconnection events in a pair plasma have shown that  highly 
         variable radio emission, with an extremely  flat spectrum ($s\sim -0.17$) is a
          natural outcome with pairs accelerated to Lorentz factors of up to 
          $\sim 70$ (e.g. Figure  2  in Jaroschek et al. 2004). 
           The flat spectrum is observed up to the cutoff frequency at  $\sim 100$ GHz (e.g. Figure 4 in Jaroschek et al. 2004).
            These simulations show that the highly dynamic non-linear
            evolution of  thin current sheets serve as the fundamental plasma
             scenario  to obtain a flat synchrotron power spectra in pair-dominated 
             environments.  The emitted synchrotron emission  is highly linearly polarized, with spectral polarization varying from $> 50\%$ around 1 GHz and up to $80\%$ or higher at higher
  frequencies.

             If pair generation  regions exist 
              in the vicinity of the lc then, when
              combined with magnetic reconnection events from
               bubble dissipation, it would offer a natural
               explanation for the observed flat spectrum.  
                This could also help account for the 
                 prevalence of emission at a particular rotation phase (see fig. 2 in Camilo et al. 2006)
                 which implies that  the co-rotating bubbles must preferentially dissipate at a particular co-rotating longitude.  It might also be the case that  there exist 
               a small non-uniformity at the light cylinder, caused  by
              feedback from pairs produced by bubble dissipation mechanism itself.

\subsection{Torques during radio emission}

As each bubble expands toward the light cylinder it exerts a torque
$\tau_{\rm Bub.} = \Omega\times  dI_{\rm Bub.}/dt  = \Omega\times (2 m_{\rm Bub.} r v_{\rm A} + \dot{m}_{\rm Bub.} r^2)$ where $I_{\rm Bub.}= m_{\rm Bub.} r^2$
 is the moment of inertia of a bubble at a radius
 $r$ from the star; here $\dot{m}_{\rm Bub.}$ is the rate of change of the bubble mass.
  For $\dot{m}_{\rm Bub.}=0$,
   the total torque exerted, $\tau = N_{\rm Bub.} \tau_{\rm Bub.}$, is then 
  \begin{equation}
  \tau  \sim  8.8\times 10^{32} \ {\rm dyn\ cm} \frac{R_{\rm in, 15}^{6}}{P_5}  t_{\rm days}^{-3/7}\ .
  \end{equation}
   The corresponding frequency derivative, $\dot{\nu}= - \tau /(2\pi I_{\rm QS})$ with
    $I_{\rm QS}\sim 10^{45}$ g cm$^2$ for the star, is 
    \begin{equation}
    \dot{\nu} \sim - 1.4\times 10^{-13} \ {\rm s}^{-2}\  \frac{R_{\rm in, 15}^{6}}{P_5 I_{\rm QS, 45}} t_{\rm days}^{-3/7}\ .
    \end{equation}
    This gives $\dot{\nu} \sim - 1.9\times 10^{-12} \ {\rm s}^{-2}$ and 
    $\dot{\nu} \sim -  10^{-13} \ {\rm s}^{-2}$ for XTE J1810$-$197 and 1E 1547.0$-$5408,
    respectively, which  is of the right  order of magnitude compared to what is observed.

     In our model, 
      the torque decreases in time (as the X-ray decreases)
       in accordance with what has been observed in the case of XTE J1810$-$197 (Camilo et al.
      2007c).     However, for 1E 1547.0$-$5408,
       the torque increased in time as the X-ray flux decreased.
        To explain the case of 1E 1547.0$-$5408 we recall that 
     the first term ($\Omega\times  2 m_{\rm Bub.} r v_{\rm A} $) in the torque equation gives a decreasing torque with radius
 while the second one ($ \Omega\times\dot{m}_{\rm Bub.} r^2$) gives an increasing torque with radius.
 Since the radius increases in time this is equivalent to the
  torque changing accordingly in time. In order for the torque
 to increase with time then $\dot{m}_{\rm Bub.} r^2 > 2m_{\rm Bub.} r v_{\rm A}$ which at a distance
   half way to the l.c. implies $\dot{m}_{\rm Bub.} /m_{\rm Bub.}  > 10^{-5}/P^{7/2}$.
 Thus a slight increase in the bubble's mass during propagation
 can give an increasing torque with time.

\section{Discussion}

 \subsection{Transient versus normal AXPs during quiescence in our model}
        \label{sec:transition}

       The two sources studied here are termed transients in the sense that in
quiescence their measured temperatures are as low as those of
 some ordinary  young neutron stars.  Only during their bursting phase
  that their bolometric luminosity becomes comparable
   to that of a typical AXP in quiescence (i.e. a bolometric
    luminosity of $\sim 10^{35}\ {\rm erg\ s}^{-1}$; see Figure 2 in OLNIII).
   From approximate outburst time, the estimated initial bolometric luminosities 
   for the two transients are $\sim
   2\times 10^{35} d_{3.3}^2\ {\rm erg\ s}^{-1}$.  
   
   In our model, to understand the difference between 
   transient and normal AXPs during quiescence, one
   should note that   the magnetic field would impose co-rotation of the ring's atmosphere (thus
                  no accretion onto the star)
                  as long as the magnetic energy density anywhere along
                   the ring's atmosphere exceeds the Keplerian energy density. The condition is
                   \begin{equation}
                   \label{eq:zeta}
                 \zeta=\frac{\rho_{\rm atm}v_{\rm K}^2}{B^2/8\pi} \simeq  0.2 \frac{T_{\rm keV}^{3/2}M_{1.4}R_{\rm out, 25}^5}{B_{\rm s, 14}^2 R_{\rm QS, 10}^6} < 1\ .
                 \end{equation}
                 
                   Thus  during the evolution  of the source as long as the condition above is satisfied
                     there will be no accretion and the object's emission is dominated by 
                     vortex expulsion.

  The  star's magnetic field decays at a rate (see \S 3 in Niebergal, Ouyed, \& Leahy 2006)
                 \begin{equation}
                 B_{\rm s} = B_0 \left(1+\frac{t}{T}\right)^{1/6} \,
                 \end{equation}
                 where $T =  840\ {\rm s}$ is the characteristic age due to vortex
                  expulsion using our fiducial value of $B_{\rm QS,0}=10^{15}$ G
                   and $P_{\rm QS,0}=1$ ms for the star's surface magnetic field and period
                    at birth, respectively.  Simultaneously, 
                   the ring  spreads viscously outward (i.e. radially)  at a rate given 
                     by eq.(A.7) in OLNII 
                     \begin{equation}
                     (\Delta r)_{\rm ring} \simeq 7.8\ {\rm km} T_{\rm keV}^{5/4} t_{\rm yr}^{1/2}\ ,
                     \end{equation}
                     where  time is in years.
                     The ring's radius at birth is given by equation (2) in OLNII and
                      is $R_{\rm out,0} \simeq 15\ {\rm km}/m_{-7}^2$ where $m_{-7}$ is
                       the ring's mass is in units of $10^{-7}M_{\odot}$ so that it spreads from
                        $R_{\rm out, 0}$ to $R_{\rm out}=R_{\rm out, 0}+  (\Delta r)_{\rm ring}$; clearly for sources
                         thousands of years old (as is the case here) the ring's outer
                         radius is simply given by $R_{\rm out}\sim (\Delta r)_{\rm ring}$. 
                         The condition for no accretion, $\zeta <1$, is then equivalent to 
                   \begin{equation}
                   t < 1300\ {\rm yrs} \times T_{\rm keV}^{-93/26}\times B_{0,15}^{12/13}\ .
                   \end{equation}
                   To get equation (24) we start with condition (21) and  $R_{\rm out}=\Delta r_{\rm ring}$
 (i.e. eq. 23) and replaced $B_{\rm s}$ from eq. (22) with $t >> T$;  The QS mass and radius
 are kept to their fiducial values. This implies that for $B_{0}=10^{15}$ G and  $0.2 < T ({\rm keV}) < 0.4$,
     the transient age is $3\times 10^4 < t ({\rm yrs}) < 4\times 10^5$.
                     Transient AXPs, we speculate, 
                      should be younger than accretion-dominated
                        sources (such as 1E2259$+$586 and 4U0142$+$61) in our model.

                       The magnetic field decay from vortex
               expulsion implies that magnetic field lines deflate radially inwards\footnote{Inside the star, the field
               lines expand outwards following the vortices.}, so we expect that violation
                of co-rotation to first occur at the ring's outer edge.                         
                      Also, at the outer edge of the disk, there is magnetic field of opposite $B_{\phi}$ being produced when matter drags field lines.
 This is an unstable situation since reconnection can occur between adjacent field lines of opposite polarity leading to accretion along the reconnecting field
 lines.     Anywhere else on the surface of the disk, whenever $\zeta >1$, the atmosphere winds up the field
            lines  without reconnecting them until the magnitude of the magnetic field is strong enough to re-inforce corotation,
 ie. the magnetic field stiffens, leading to a stable balance.

         In summary, a transient  AXP quiescent phase is dominated by  X-ray
          emission from vortex expulsion only ($L_{\rm X,vortex}\simeq 2\times 10^{34}\dot{P}_{-11}$ erg s$^{-1}$)  with the source continuing to evolve along the vortex band (see Figure 2 in OLNIII).
           On the other hand,  a  typical AXP (older ring-bearing source) quiescent phase
                        is dominated by emission from the 
           HS induced by accretion from the outer edge of the ring with $L_{\rm acc.}$ given by eq.(\ref{eq:hslum}).
           These will evolve horizontally (at constant $L_{\rm HS}\sim 10^{35}$ erg s$^{-1}$)  as discussed
                  in OLNII (see  Figure 2 in OLNIII).     
                  
                  It is only during the bursting phase that transient and regular AXPs
                   would look the same since they both experience the BF effect
                    and the related accretion and feedback process. In the
                     transient case, as the BF dies out when reaching the outer edge,  
                      the system becomes dominated again by vortex expulsion
                       while regular AXPs resume their accretion dominated (from
                        the outer edge) quiescent phase.

 \subsection{Birthrate}

In our model, a QS-ring system experiences an X-ray/radio outburst every few hundred years.
    Since we have observed 2 in  a few years, located at distances
    of 3 to 9 kpc away (meaning  we see about 30\% of them), it is
    suggestive of a rate of 1 per year for the whole galaxy. This implies
     a total population of $(1/{\rm yr})\times 100\ {\rm yrs}\sim 100$ in the galaxy. Since the
      ring will be consumed on a timescale of $\sim  10^4$-$10^{5}$ years (see
      eq. 5 in OLNIII) this gives a  birthrate of  (1 per 1000 years) to (1 per 100  years).     
          Within uncertainties, the birth rate
         of transient AXPs derived above is consistent   with the expected
        birthrate of AXPs ($\sim 1/300\ {\rm yrs}$; Gill\&Heyl 2007; Leahy\&Ouyed 2009).
         As we argued in the previous section, in our model transient AXPs
         evolve into typical AXPs thus sharing the same birthrate.

\subsection{Model features and predictions}

 The general predictions, starting with the X-ray emission, in our model are:

  \begin{itemize}

 \item \underline{The\ 2\ Blackbodies}:   
 
  Overall, during burst and quiescence, the cool BB (from  Keplerian ring; or from the co-rotating
     shell as in OLNI) arises from reprocessing
     radiation of the hot BB from the star (either vortex annihilation or accretion HS).
     We suggest, the relations in Figure 6 in Nakagawa et al. (2009)  can be explained
      in the context of reprocessing during both quiescence and burst.
        Table 1, summarizes the different emission components for different objects in
  different states in our model.   Compared to SGRs (see OLNI) and transient AXPs,
   typical AXPs acquire an additional BB  from the HS during quiecence.
    Only during bursting do transient AXPs acquire a HS. As for SGRs in our model,
     we recall (see OLNI) they  are born with a co-rotating shell (i.e. non-Keplerian degenerate ring;  
         see OLNI).    Future work will consider hard X-ray emission from non-thermal
          processes related to magnetic reconnection following  vortex expulsion and to accretion onto the QS.

          \item \underline{During\ X-ray\ Burst}: Since the accreted material in our model consists mostly of  dissociated 
         iron  ($Z \sim 13$), we predict some sort of signatures either
          during channeling along the field line or on impact on the HS -- maybe absorption lines or proton
          cyclotron lines ($\sim 0.14 B_{\rm QS, 14}$ keV) from any of element in the Ne-to-S group (e.g. Ne, Al, Si).
           These signatures should be common for both transient and typical AXPs
            in our model.
          
          \item     \underline{During\ X-ray\ Quiescence}:  When edge accretion occurs the 
             signatures (e.g. absorption lines or proton
          cyclotron lines)  should be from the $Z=26$ rather than the $Z \sim 13$ nuclei.

\item \underline{X-ray\ variability }:  During both the quiescent and bursting phases, 
     X-rays from the ring$+$atmosphere system 
 should be unpulsed and may carry the Keplerian sgnature via millisecond variability.

   \end{itemize}

 The general predictions for the radio emission are:

  \begin{itemize}
   
     \item   \underline{Radio\ Delay}:  As can be seen from equation (\ref{eq:radiodelay}), the shorter the period
           of the star the smaller the delay, $t_{\rm delay}$,  between the X-ray burst and the
           radio emission. Interestingly, the 2 radio emitting AXPs so far are observed
           are those with the smallest period.

       \item  \underline{Radio\ Flux}:  Furthermore, as can be seen from eq.(\ref{eq:Lradio}), as $P$ gets larger the radio
        following X-ray burst gets very faint making it more  difficult to detect. Combined with long $t_{\rm delay}$
       for large $P$, we argue these to be the reasons why radio
       is not observed following X-ray bursts in AXPs with  higher ($\sim 10$ s) period.

 \end{itemize}

  Finally, we list specific   
 predictions in our model in the case of XTE J1810$-$197: 

\begin{itemize}

\item \underline{XMM\ warm\ BB }: The XMM warm BB
  for XTE J1810$-$197 (ring$+$atmosphere system in our model) will evolve
   back to one single BB  (ROSAT BB) with temperature $T_{\rm q}$.

 \item \underline{XMM\ hard\ BB }: The XMM hard BB
   (the accretion  HS in our model) will disappear  following burst
    (i.e. once the Bohm front reaches $R_{\rm out}$).  However, 
   we expect edge effects    
   to appear  (e.g.  flattening or even jump in the hot BB area; see last panel in Figure 3)  just before
    accretion shuts off.

  \end{itemize}

 \begin{table}[t!]\label{tab:components}       
\caption{Thermal components in our model}
\centering
\begin{tabular}{|clc|c|}
\hline
Sources &  Quiescent phase  & Bursting phase \\\hline
SGRs  &      2 BBs  &  3 BBs \\
Transient AXPs &   2 BBs  &  3 BBs \\
Typical AXPs &   3 BBs &  3 BBs \\
\hline
\end{tabular}\\
\end{table}

\subsection{Further implications}

                 There exist two aspects of our model that might
          provide some answers to fundamental issues in pulsar 
magnetospheres\footnote{A disk model for X-ray emission
     from pulsars was considered by Michel \& Dessler (1981). 
However, their disk is fundamentally different from our ring in that,
where they hypothesize electron degenerate material (left over from a supernova),
in our model the material is completely relativistic-degenerate (from a Quark-Nova).
This difference has many consequences, the most notable are a more efficient accretion
mechanism (conversion energy from the hadron to quark transition as well as gravitational
energy is released), and a slower (viscous) spreading rate for the ring.}.
           First,   the source in the quark-nova model is born as an aligned rotator
            and  secondly is the fact that here the degenerate  ring,
          by Keplerian shear, is a natural source of plasma (carried
           by the bubbles to the l.c.) for the magnetosphere.  
           Also, pairs are naturally supplied to the magnetosphere 
            by vortex annihilation. 
           This is left as an avenue for future investigation.

 \subsection{Model Limitations} 
 \label{sec:limit} 
 
Our model suffers from a  few caveats : 
{\bf (i)} The interaction between the BF and the magnetic field in the
atmosphere is rudimentary at this stage. 
 Understanding the exact mechanism of feeding material from degenerate
 ring into the atmosphere and onto the field lines within the Bohm front 
  is essential as this process is behind the feedback process
   between the ring and the HS on the star.  This is a complex
    problem/system that would require detailed MHD simulations
       before we can confirm this aspect of our model;          
{\bf  (ii)} While the bubble generation mechanism we propose  is common
  in systems involving dipole files threaded by a disk
   on one end and a star on the other, the outward propagation of the
    bubbles within the l.c. depends crucially on conditions
     within the magnetosphere.  In particular, the value $\alpha=1$  was chosen solely 
   on the assumption that the bubbles
    will propagate in an equatorial wind (given the Keplerian nature of the bubbles at birth)
     within the magnetosphere. For highly magnetized, aligned
     rotators, the physics of the magnetosphere within the l.c.
      is yet to be understood and the problem solved.  Furthermore, 
      the bubble dissipation mechanism as they reach the l.c.  remains
 to be demonstrated. The detailed structure of the thin shear layer at the light cylinder is not well  
  studied and would require numerical simulations. For now our explanation
   is based on the assumption that the   thin 
 transition shear layer  at the l.c. slowly destroys the bubble by shearing off the part that touched the light cylinder.

  \section{Conclusion}
  
  There are two fundamental components in our model for
   AXPs and transient AXPs namely, the QS and the Keplerian ring.
   In quiescence, vortex annihilation on the QS gives rise to thermal
    and non-thermal X-ray emission. The ring reprocesses the emission
    to give a second cooler BB emission component. Outburst
    is triggered by accretion of a small inner part of the ring (i.e. the wall).
     The two main consequences are production of light ($Z \sim 13$)
      nuclei and triggering MHD accretion onto the QS (yielding the HS).
  The interplay between the Bohm diffusion (i.e. $R_{\rm BF}$ term) and depletion of light nuclei
   (i.e. $\mu$)   gives rise to a rich behavior, necessary
        in order to  account for the observed   behavior of XTE J1810$-$197. 
         Finally, one can ask if such a small Keplerian degenerate iron-rich ring  could
         form around a neutron star. Ring formation
          when the neutron star is born 
         appears implausible since a proto-neutron star  is large
          compared to the ring size. After formation, there is no obvious
          mechanism to eject degenerate material unless a violent change
          of state, like a QN occurs.


\begin{acknowledgements}
This research is supported by grants from the Natural Science and
Engineering Research Council of Canada (NSERC). We thank
 the referee for  comments that helped improve this paper.
\end{acknowledgements}


\begin{appendix}

   \section{The Bohm diffusion front}\label{app:BF}
   
   The hot front propagates from the inner parts of the ring at a Bohm diffusion rate given
    by 
    \begin{equation}
    \Delta r  = \psi_{\rm BF} \left(6.25\times 10^{-4} \frac{T_{\rm kev}}{B_{\perp, 13}}\right)^{1/2} t^{1/2}\ ,
    \label{eq:bohm}
    \end{equation}
     where $B_{\rm \perp, 13}$ is the magnetic field component perpendicular
      to the front motion in units of $10^{13}$ G.  The parameter $\psi_{\rm BF}$
       carries some uncertainty related to the fact that the above formula is
       semiempirical  (i.e. the constant 1/16 in the original Bohm diffusion expression has
        no theoretical justification; see \S 5.10 in Chen 1984).  This parameter  is derived self-consistently in
       our model as explained in  \S \ref{sec:tau13} below.  For a 
 qualitative derivation of Bohm diffusion we refer the interested reader
  to  Alexeff\&Rader  (1991) and Belyaev (2001).

      The ring geometry allows us to write $B_{\perp, 13} \simeq \tan{(\alpha_c)} B_{\rm BF, 13}$ where to a first approximation $\tan{(\alpha_c)} = (H_{\rm atm., out}^{z}- H_{\rm atm., in}^{z})/(R_{\rm out}-R_{\rm in})$ and  $B_{\rm BF} \simeq B_{\rm av.} = B_{\rm QS} (R_{\rm QS}/R_{\rm av.})^3$
       where $R_{\rm av.}=(R_{\rm in}+R_{\rm out})/2$; here $B_{\rm QS}=\sqrt{3\kappa P \dot{P}}$
       and $R_{\rm QS}$ are the star's magnetic field and radius, respectively. The temperature cancels out from eq.(\ref{eq:bohm}) since $B_{\perp}\propto
             H_{\rm atm.}^{z}\propto T_{\rm keV}/\mu$ so that,  
          \begin{equation}
          \label{eq:a2}
           \Delta r  \sim 10^{-2}\ {\rm km}\ \psi_{\rm BF} \left( \frac{\mu_{3.3} }{B_{\rm av., 13}R_{\rm out, 15}^{1/2}}\right)^{1/2} t_{\rm days}^{1/2}\ .
          \end{equation}


           \section{Centrifugal ejection and channeled accretion}\label{app:MHD}

           As the system (ring$+$atmosphere) is heated,
       the boundary between the non-degenerate atmosphere and the degenerate
       $Z \sim 13$ layer  moves downwards into higher density layers (see eq.(\ref{eq:atmosphere})).
         Thus Keplerian ring material is fed into the atmosphere 
          and  is ejected as a magnetohydrodynamic (MHD) wind as outlined below (see also
          lower panel in Figure 2).

           At keV temperatures the gas is sufficiently ionized 
           everywhere in the atmosphere that ideal MHD can be used.
              Since the Lorentz force only has components perpendicular to the field,
            the gas is free to move along the co-rotating field line under the influence
             of other forces.
             Under these conditions, it
              has been shown that the wind can be launched centrifugally if the 
              field direction is inclined at an angle less than $60^o$ to the radial
               direction (Blandford\&Payne 1982).
               Using conservation of specific angular momentum ($\propto r^2 \Omega$),
            a non-degenerate Keplerian particle (i.e. $\Omega_{\rm K}$)  loaded at a  
            footpoint $r_0$ will be flung out to larger radii;  the so-called ``bead-on-wire" analogy.      
               These conditions are easily met in our model so that
                the newly unveiled Keplerian material (following heating of the
                degenerate ring) finds itself threaded by the
            highly inclined  magnetic field and is flung out centrifugally.
             The wind is then channeled towards the star's surface by the
             strong dipole.

               The mass flux, $\dot{m}_{\rm wind}$, 
              is regulated by conditions (i.e. density) at the slow mode
               critical point:
               \begin{equation}
               \label{eq:b1}
               \dot{m}_{\rm acc} = 4\pi R_{\rm BF} \zeta_{\rm sp} (\rho_{\rm atm.,b}H_{\rm atm.,b}^{z}- \rho_{\rm atm.,q}H_{\rm atm.,q}^{z})
               v_{\rm atm, th}\ ,
               \end{equation}               
                where $(\rho_{\rm atm.,b}H_{\rm atm.,b}- \rho_{\rm atm.,q}H_{\rm atm.,q})= \rho_{\rm atm.,b}H_{\rm atm.,b}\times f_{\rm K}(T_{\rm b})$ describes the mass injected from the Keplerian ring
     into the  atmosphere.  
               The function   $f_{\rm K}(T_{\rm b}) = 1-  T_{\rm q}^{5/2}/T_{\rm b}^{5/2}$  
       evolves from $\sim 0.8$ to zero as the non-degenerate Keplerian material is depleted  
       and the system evolves back to its quiescent phase.  In a sense it acts as a ``valve" that 
          shuts-off accretion after the heating front has passed.
          
                The accretion rate, $\dot{m}_{\rm acc.}\sim \dot{m}_{\rm wind}$,  can be recast to
    \begin{equation}
    \label{eq:mdotacc}
     \dot{m}_{\rm acc.}  = 1.77\times 10^{17} \zeta_{\rm sp, 0.01} \frac{T_{\rm  b, keV}^3 f_{\rm K}(T_{\rm b}) R_{\rm BF, 15}^{5/2} }{\mu^{3/2}}\ .
     \end{equation}
                  The factor $\zeta_{\rm sp}$ carries the uncertainty on the exact
                   location of the slow MHD mode. We will adopt $\zeta_{\rm sp} \sim 0.01$ as inferred
                    from 2-dimensional (Ouyed et al. 1997) and 3-dimensional (Ouyed et al. 2003) MHD simulations of disk winds.                     
                     The slow MHD point governs the mass loss rate.
     The degenerate ring responds by suppling  mass at a rate $\dot{m}_{\rm BF}\sim \rho_{\rm atm} 2\pi
                   R_{\rm BF} d_{\rm BF}v_{\rm \perp}$ where $v_{\rm \perp}$ is the velocity
                   of degenerate Keplerian material crossing the boundary to non-degeneracy\footnote{ We recall that in the degenerate Keplerian disk,
              lower viscosity and conservation of angular momentum imply
                mass flows in the ring    is along lines of constant angular momentum
                 which is nearly vertical (i.e. $z$-direction; see Appendix A in OLNII).   
                  Thus as the non-degenerate Keplerian material is ``sucked up"
                  in the wind, more material is supplied almost vertically from the underlying Keplerian material.}, and 
                    $d_{\rm BF}$   is the radial width of the BF as illustrated in Figure 2.

   \subsection{Accretion shut-off}

                      The accretion relies upon new Keplerian  material being fed from the
 degenerate ring to the atmosphere in the BF region, and ends once
 the magnetic field has re-inforced co-rotation of the atmosphere.
 This defines the inner edge of the BF as illustrated in the lower
 panel of figure 2. 
                      The penetration timescale
                       can be estimated from equation (16)
                        in OLNII to be   $\tau_{\rm B, K}\sim 4~ {\rm days}\ H_{\rm K, cm}^2/T_{\rm keV}^{1/2}$
                       where the depth of the non-degenerate Keplerian layer
                        $H_{\rm K}\sim 10^{-3} H_{13}\sim 10^{-6} H_{\rm ring}$ is  in units
                        of centimeters.
                        The size (i.e. radial width) of the BF front is found from  
                   $d_{\rm BF}= v_{\rm BF}   \tau_{\rm B, K} \sim 10^3$ cm which is of the same order as the
                    depth of the $Z \sim 13$ layer.

    \subsection{Depletion timescale of the $Z \sim 13$ nuclei}  
    \label{sec:tau13}

            We recall that following wall accretion and subsequent
            irradiation of the ring,  up to  $10^{46}$ of iron nuclei are dissociated forming
            the $Z \sim 13$ layer.
    Subsequent depletion of the $Z \sim 13$ nuclei by accretion   leads
    to a return to an iron-rich atmosphere  (i.e. $\mu_{\rm q}\sim 3.3$) as shown
     in OLNII:
    \begin{equation}
   \frac{1}{\mu} =  \frac{1}{\mu_{\rm q}} + (\frac{1}{\mu_{\rm b}}-\frac{1}{\mu_{\rm q}})\exp^{-t/\tau_{13}}\ ,
   \end{equation}
    where $\tau_{13}$ is the depletion timescale  given by $\tau_{13}\sim m_{13}/\dot{m}_{\rm acc.}$
                where  $m_{13}\sim  N_{13}~\times (13~ m_{\rm H})$.
                        For typical
            values  following wall accretion $\mu =\mu_{\rm b}  \sim 2.3$, $R_{\rm BF}\sim R_{\rm in}$,
             $Y_{\rm BF}\sim 0.4$, and $\eta=0.1$, we find             
             \begin{equation}
             \label{eq:tau13}
          \tau_{13}\sim  1.4\times 10^{4}\ {\rm days}\ 
          \frac{N_{13, 46}}{\zeta_{\rm sp, 0.01}^4}\ ,
          \end{equation}
           where $N_{13, 46}$ is the total number of dissociated iron nuclei  in units of $10^{46}$.
           
            We have already estimated the size of the BF region
                         above which we found to be of the same order as the depth
                          of the  $Z \sim 13$ layer
                      which implies that our scenario self-consistently leads to the accretion
                       of most of the $Z \sim 13$ material during burst (i.e. during the time it takes
            the BF to comb the ring's surface).   In other words, the 
            depletion timescale can also be estimated from eq.(\ref{eq:a2}) by taking $\Delta r = (R_{\rm out}-
             R_{\rm in})$ and $\mu = \mu_{\rm b}\sim 2.3$ to get
             \begin{equation}
             \tau_{\rm 13} \sim \frac{1.5\times 10^4}{\psi_{\rm BF}^2}\ {\rm days}\ (R_{\rm out}-R_{\rm in})_{\rm km}^2 R_{\rm out, 15}^{1/2} B_{\rm av., 13}\ ,
             \end{equation}
              where $(R_{\rm out}-R_{\rm in})$ is in units of kilometers. 
                         Equating the previous two equations we find
          \begin{equation}
          \psi_{\rm BF} \sim \frac{\zeta_{\rm sp, 0.01}^2 (R_{\rm out}-R_{\rm in})_{\rm km} R_{\rm out, 15}^{1/4}
          B_{\rm av.,13}^{1/2}}{N_{13,46}^{1/2}}\ ,
          \end{equation}
           which fine tunes the speed of the BF in our model (i.e. eq.(\ref{eq:a2})).

\section{Ring-Pole interaction: the hot spot}\label{app:HS}

As illustrated in Figure 2, there is a direct link between the BF 
 moving outward along the ring surface (increasing its area)
  while decreasing    the area of the HS on the surface of the star.
    It is straightforward to show that the total HS area (both poles) is
    \begin{equation}
    \label{eq:hsarea}
    A_{\rm HS} = 4\pi R_{\rm QS}^2 \times (\cos{\beta_{\rm BF}} - \cos{\beta_{\rm in}})\ ,
    \end{equation}
     where the angle $\beta$ defines the colatitude (i.e. measured from the polar  axis) on the star's surface where
      material channeled along a given field line lands . We approximate the field
       by  a dipole configuration, giving us
       \begin{equation}
       \sin{\beta} = \left( \frac{R_{\rm QS}}{R}\right)^{1/2} \sin{\alpha}\ ,
       \end{equation}
       where
       \begin{equation}
        \alpha = \frac{\pi}{2} - \arctan{\frac{H_{\rm ring}}{R}}\ ,
       \end{equation}
        is the angle from the polar axis to the footpoint, at radial distance $R$,  
        of the field line  where the Keplerian material is loaded; $H_{\rm ring}\sim 2.68\  {\rm km}\rho_{\rm ring, 9}^{1/6} R_{15}^{3/2}$ is the ring's vertical height as given in \S \ref{sec:ring}. For 
        our fiducial values, $R_{\rm in} = 15$ km, $R_{\rm out}= 25$ km
          and $R_{\rm QS}= 10$ km, we find $(\cos{\beta_{\rm out}} - \cos{\beta_{\rm in}})\sim 0.192$
           which gives an initial area of the HS at one pole of the order of $\sim 120$ km$^{2}$.
      
\end{appendix}


\end{document}